\documentclass[twocolumn,secnumarabic,amssymb, nobibnotes, aps, prd,superscriptaddress]{revtex4-2}

\usepackage{amssymb}
\usepackage{amsmath}
\usepackage{amsthm}
\usepackage{mathtools}
\usepackage[font=small,labelfont=bf,singlelinecheck=off]{caption}

\usepackage{graphicx}
\usepackage{tikz}
\usepackage{float}
\usepackage{subfig}
\usepackage{enumitem}
\usepackage[normalem]{ulem}
\usepackage{multirow}
\usepackage{rotating}

\makeatletter
\DeclareFontFamily{U}{tipa}{}
\DeclareFontShape{U}{tipa}{m}{n}{<->tipa10}{}
\newcommand{\arc@char}{{\usefont{U}{tipa}{m}{n}\symbol{62}}}%

\newcommand{\arc}[1]{\mathpalette\arc@arc{#1}}

\newcommand{\arc@arc}[2]{%
  \sbox0{$\m@th#1#2$}%
  \vbox{
    \hbox{\resizebox{\wd0}{\height}{\arc@char}}
    \nointerlineskip
    \box0
  }%
}
\makeatother

\newcommand{\cosp}[1]{\cos\left(#1\right)}
\newcommand{\arccosp}[1]{\arccos\left(#1\right)}

\newcommand{\ket}[1]{\left| #1 \right\rangle}
\newcommand{\bra}[1]{\left \langle #1 \right |}
\newcommand{\ketbra}[1]{\left| #1 \right\rangle\left \langle #1 \right |}

\newcommand{\braket}[2]{\langle #1 | #2 \rangle }
\newcommand{\ketsmall}[1]{| #1 \rangle}

\DeclarePairedDelimiter\ceil{\lceil}{\rceil}

\begin{document}

\title{Quantum algorithm for credit valuation adjustments}%

\author{Javier Alcazar}%
\affiliation{Zapata Computing, Inc.}

\author{Andrea Cadarso}%
\affiliation{BBVA Corporate \& Investment Banking, Calle Sauceda 28, 28050 Madrid, Spain}

\author{Amara Katabarwa}%
\affiliation{Zapata Computing, Inc.}

\author{Marta Mauri}%
\affiliation{Zapata Computing, Inc.}

\author{Borja Peropadre}%
\affiliation{Zapata Computing, Inc.}

\author{Guoming Wang}%
\affiliation{Zapata Computing, Inc.}

\author{Yudong Cao}%
\email{yudong@zapatacomputing.com}
\affiliation{Zapata Computing, Inc.}


\begin{abstract} 
Quantum mechanics is well known to accelerate statistical sampling processes over classical techniques. In quantitative finance, statistical samplings arise broadly in many use cases. Here we focus on a particular one of such use cases, credit valuation adjustment (CVA), and identify opportunities and challenges towards quantum advantage for practical instances. To improve the depths of quantum circuits for solving such problem,  we draw on various heuristics that indicate the potential for significant improvement over well-known techniques such as reversible logical circuit synthesis. In minimizing the resource requirements for amplitude amplification while maximizing the speedup gained from the quantum coherence of a noisy device, we adopt a recently developed Bayesian variant of quantum amplitude estimation using engineered likelihood functions (ELF). We perform numerical analyses to characterize the prospect of quantum speedup in concrete CVA instances over classical Monte Carlo simulations. 
\end{abstract}

\maketitle

\section{Introduction}

\begin{figure*}
    \centering
     \includegraphics[scale=0.35]{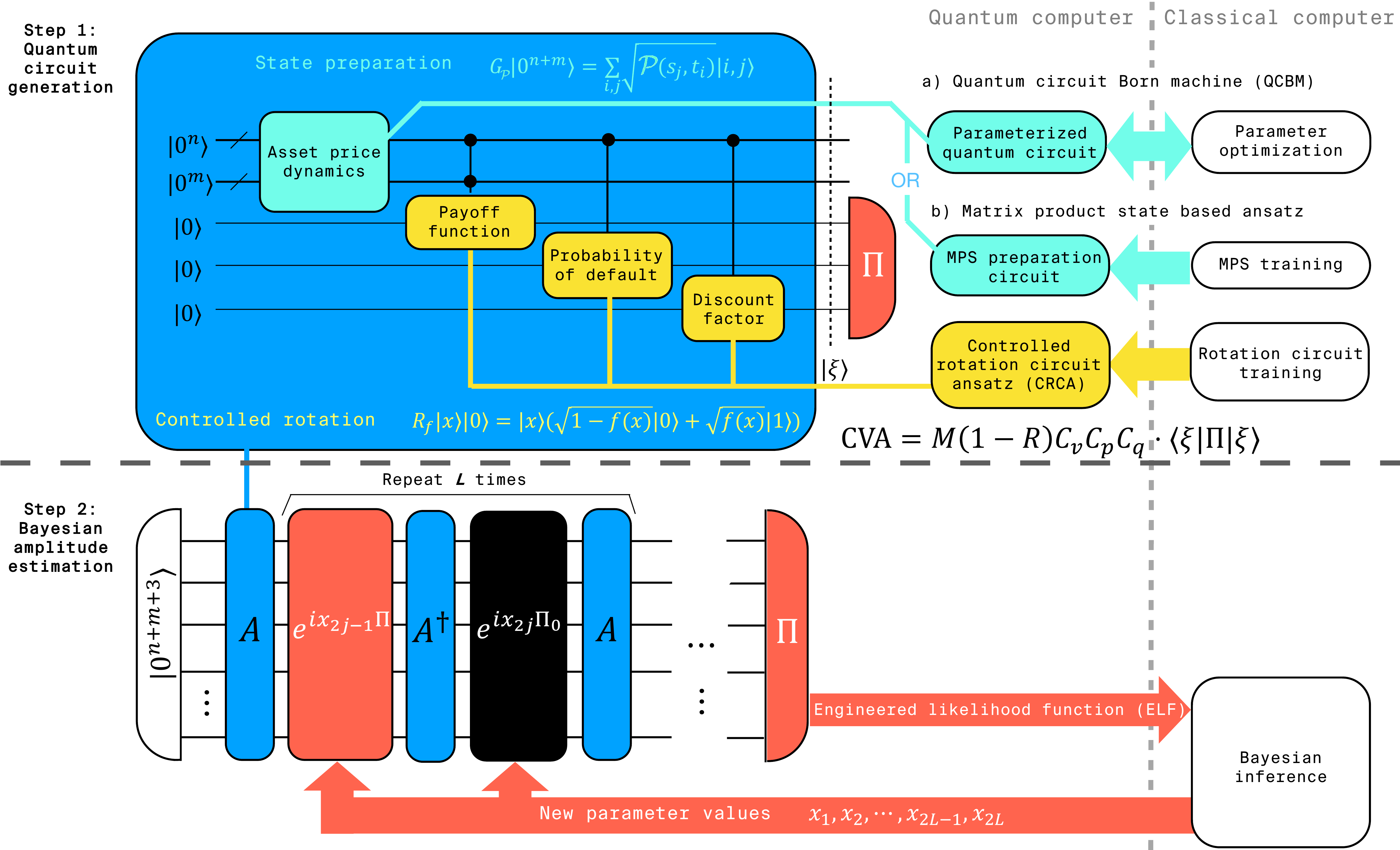}

    \caption{Overview of the quantum approach to credit valuation adjustment proposed in this work. We start by inspecting the components that make up the CVA quantity (Section \ref{sec:cva}), which translate to the structure of the quantum circuit $A$ in Step 1. It consists of state preparation and controlled rotations (Section \ref{sec:Map_to_quantum}). We expand on the quantum circuit construction in Section \ref{sec:quantum_circuit}, where we investigate two alternatives to the state preparation subroutine: quantum circuit Born machine (QCBM) in Section \ref{subsec:QCBM} and matrix product states (MPS) in section \ref{subsec:MPS}. For controlled rotation, we propose controlled rotation circuit ansatz (CRCA) in Section \ref{sec:crca_section}. For both MPS and CRCA, training occurs only on the classical computer and the resulting parameters are used for constructing the quantum circuit. For QCBM, one trains the quantum circuit iteratively between the quantum and classical computer. At this point the problem of estimating CVA is reduced to estimating the expectation of an observable $\Pi$ with respect to the output state $|\xi\rangle=A|0^{n+m+3}\rangle$ of the quantum circuit $A$. We then move on to Step 2 to perform the amplitude estimation using engineered likelihood functions (Appendix \ref{sec:ELF}).}
    \label{fig:cva_concept_figure}
\end{figure*}

Statistical simulation tasks are often the most computationally expensive exercises that banks perform. One important class of such exercises is counterparty risk analysis, which has gained increasing importance in the recent years since the Great Recession in the late 2000s. In the aforementioned financial crisis, banks lost tremendous amount of capital in counterparty credit default during derivative transactions, which led to specific regulations and capital requirements. Therefore, risk analyses need to be in place to calculate the precise amounts by which the prices of the derivatives should be adjusted to hedge against the risk of counterparty default, and fall under the general term of credit valuation adjustments (CVA). In response to the credit losses during the Great Recession, the Basel Committee for Banking Supervision (BCBS) has defined regulatory requirements for CVA calculations \cite[50.31-50.36]{BCBS20}. The regulation demands that CVA be calculated by simulating the stochastic paths of the underlying exposures, which is frequently carried out by Monte Carlo methods. The probabilistic nature of this process means there will be inherent statistical errors in the resulting CVA estimation. In order to suppress such statistical errors, one must increase the number of stochastic paths sampled. Therefore, a typical CVA calculation for a derivative product involves calculating statistical averages over a large number of price trajectories of the underlying asset(s) of the derivative as well as possible default scenarios of the counterparty. A rough estimation \cite{CVA} shows that a large number of Monte Carlo samples is needed for meeting the standards laid out in Basel Committee on Banking Supervision's July 2015 consultation document regarding CVA calculations.

The stochastic simulation of the prices of the underlying assets as they change over time is one of the key ingredients of most CVA calculations, except for cases where the expected exposure can be computed analytically. The price fluctuation is typically described as a stochastic process, which is defined for every point in time. Simulating such a stochastic process on modern digital computers introduces a discretized time grid, which introduces a discretization error $\epsilon_D$. Each simulation generates a concrete path of how the price(s) of the asset(s) change(s) over time, which supplies one sample in the Monte Carlo simulation for estimating the expected payoff of the financial instrument that is based on the asset(s). The general goal of the Monte Carlo simulations in financial use cases such as CVA can be described as estimating the expectation value $\mathbb{E}[f(S)]$ of some function $f$ of a set of random variables $S$. Since one is only able to draw a finite amount of samples or paths on a computer, there is a statistical error $\epsilon_S$ associated with each simulation. To estimate the expectation within additive error $\epsilon_S$, generally $\mathcal{O}(1/\epsilon_S^2)$ samples are required. On a classical computer, discretization error introduces additional cost in the simulation. For a discrete approximation scheme of order $r$ on a grid of time interval $\Delta t$, the discretization error is $\epsilon_D=\mathcal{O}(\Delta t^r)$ \cite{Kloeden1992}. This gives rise to an extra factor of $\mathcal{O}(1/\epsilon_D^{1/r})$ in the cost of classical simulation. However, such overhead factor can in practice be either mitigated by adopting higher order approximation schemes or multi-level schemes \cite{Giles2015}. The cost scaling of $\mathcal{O}(1/\epsilon_S^2)$ with respect to statistical error $\epsilon_S$ is a more fundamental artifact of statistics that is in general hard to overcome classically.


The advent of quantum computing presents an exciting opportunity to break the barrier of the classical cost scaling with respect to statistical error. A typical strategy for addressing the problem of estimating $\mathbb{E}[f(S)]$ is by casting it as a problem of estimating an operator expectation: $\mathbb{E}[f(S)]=c\langle\psi|O|\psi\rangle$ for some constant $c$, state $|\psi\rangle$ and operator $O$. The problem of estimating $\langle\psi|O|\psi\rangle$ can be addressed by quantum amplitude estimation, with which early proposals \cite{amplitude_amp,quant-ph/9908083} demonstrated that one could improve the fundamental cost scaling of parameter estimation from $\mathcal{O}(1/\epsilon_S^2)$ to $\mathcal{O}(1/\epsilon_S)$. This is a significant improvement for applications requiring some control over the statistical error.
However, the quantum advantage is realized at a cost of running deep circuits of depth $\mathcal{O}(1/\epsilon_S)$ on a quantum computer. This renders the early algorithms infeasible for near-term quantum devices, which can execute circuits of only finite depth. Recently, there have been various proposals \cite{Wang2019,2006.09350,Suzuki2020,Zeng20} for realizing amplitude estimation with reduced depth circuits, at a cost of less quantum speedup compared with the quadratic advantage given by fault-tolerant quantum computers. A general goal of these proposals is to derive as much asymptotic speedup as possible by using the quantum resource available on a given quantum device, even though the speedup may be less than quadratic in many cases.

In this study, we adopt the framework of engineered likelihood function (ELF) proposed in \cite{2006.09350, 2006.09349} for carrying out the CVA calculation on a quantum computer. This is the first proposal of a quantum algorithm able to tackle the CVA problem, with in-depth discussions about concrete implementations in terms of elementary operations that can be carried out on quantum computers. The same approach can be extended to other risk analysis use cases in quantitative finance. A particular advantage of the ELF framework is that it does not impose \emph{a priori} an amount of quantum resource required for carrying out a certain task, but instead it adaptively takes advantage of however much quantum resource one can afford on a noisy quantum device.
To illustrate the concreteness we are able to associate to our solution of the CVA problem, our numerical results indicate that on a quantum computer that has all-to-all qubit connectivity, physical gate error rate $10^{-3}$, and uses the surface code of distance $18$ for error correction (with cycle time $1 \mu s$), performing CVA calculation under the specification listed in Table \ref{table:CVA_parameters} within relative error of $0.001\%$ takes around $4.9\times 10^5$ seconds, while the same calculation takes around $3.7\times 10^6$ seconds on a single-core classical computer (4GB RAM, up to 3.1GHz). 
Such comparison is certainly subject to changes in the details of both the classical and quantum implementations. However, the level of specificity at which we are able to carry out the resource estimation makes it straightforward to account for such implementation details if necessary. With the engineered likelihood function (ELF) technique developed previously \cite{2006.09350, 2006.09349} we are able to produce concrete estimates of quantum runtime as the noise and error parameters of the hardware improve.

Our overall approach for the quantum calculation of CVA can be divided into two steps (Figure \ref{fig:cva_concept_figure}): quantum circuit generation and amplitude estimation. The latter has been discussed in the preceding paragraphs. The former can be further divided into state preparation and controlled rotation subroutines. For both subroutines we have identified opportunities to drastically reduce the depths of quantum circuits compared with those produced from reversible logic synthesis (RLS) \cite{quant-ph/0208112,10.1145/780542.780546,0801.0342,10.1145/2493252.2493256}.

$\quad$\\
\noindent{\bf Related works.} There are many applications in quantitative finance such as derivative pricing and risk analysis that amount to performing integration on domains of stochastic variables. During the early days of quantum computing \cite{Nayak98,Abrams99,Novak2001}, it has been recognized that, for those integration tasks, one is able to glean a quadratic speedup in cost scaling with respect to the statistical error $\epsilon_S$ by using insights on quantum counting \cite{Brassard1998}, which later on was developed into quantum amplitude estimation \cite{amplitude_amp}. This line of inquiry was later extended to concrete quantitative finance use cases such as option pricing \cite{PhysRevA.98.022321,Stamatopoulos2020optionpricingusing} and risk analysis \cite{Woerner2019}. However, no use cases related to valuation adjustment (XVA) have been considered so far, making our proposal the first of such studies. There has been also a line of research \cite{Montanaro2015, An20} focusing on using quantum computers for accelerating Monte Carlo calculations; these studies are based on the availability of quantum oracles that can implement certain functions, without considering how these oracles are implemented. Here in our work, however, we will discuss the detailed implementation of the oracles relevant for the CVA problem using elementary quantum operations. Our broader motivation is to make the quantum algorithm as concrete as possible so that it will be amenable for comparison with existing classical solutions as well as implementation on near-term quantum devices.

\section{Credit valuation adjustment}\label{sec:cva} 
Financial derivatives are essentially contracts between two parties. 
For example, an option contract is a guarantee that one can buy or sell a set of underlying assets at a particular price before or on a specific date (depending on the type of option contract). However, it is possible that when the option contract is exercised, namely when one of the parties decides to go through with the transaction (buying or selling), 
the counterparty is not able to honor the contract, e.g., the party responsible for buying the asset(s) does not have enough capital to make the purchase. This is a default event that leads to a loss on the selling party. Since an option contract offers insurance against price fluctuation of the underlying asset(s), it has an intrinsic value for which one must pay a premium in order to enter it. In the event of default, the premium paid by the party entering the option contract is essentially lost. Therefore, if there is risk that such default may happen, the premium to be paid for the option contract should be accounted for in its price. The fair amount with which one should make the discount is the value of CVA. 

In this study, we focus on CVA problems for European Call options. In general, the CVA quantity is built from the following components:
\begin{enumerate}
    \item The probability distribution of asset prices $P(s|t)$ at time $t$. Classically, for a given future time $t=\tau$, sampling from $P(s|t=\tau)$ is typically achieved by performing stochastic simulation of how the asset price $s$ fluctuates as a function of time up to $\tau$, and the set of final price values is the set of samples from $P(s|t=\tau)$. 
    One of the most common stochastic processes used for modeling price fluctuation is geometric Brownian motion. 
    \item The net amount $v(s,t)$ gained by the purchaser of the option contract, or \emph{payoff}, at asset price $s$ and time $t$. For a given future time $t=\tau$, the \emph{expected exposure} $E(\tau)=\mathbb{E}_{P(s|\tau)}[v(s,\tau)]$ characterizes the average worth of the option at time $\tau$. Classically, this is estimated by averaging over the values of $v$ computed for each of the samples generated in the previous step. In particular, for the case of European options the payoff is the maximum value between zero and the difference between the price of the asset at maturity and a fixed price $K$ predetermined at the start of the contract, the \emph{strike price}.
    \item The probability of default $q(t)$ at time $t$. One method for modeling $q(t)$ is to consider it as a Poisson process where the Poisson parameter is time dependent. Its exact time dependence can be bootstrapped efficiently, and calibrated from market quantities such as CDS spreads \cite{Marchioro09}.
    \item The discount factor $p(t)$. It expresses the time value of money, and it is used to determine the present value of an asset in the future. The formula for the discount factor will depend on the number of periods considered for interest rate payments, where a typical choice is as continuous compound interest, which corresponds to discount factor $p(t)=e^{-rt}$ for an interest rate $r$. The interest rate can also be time dependent.
\end{enumerate}
The CVA is then calculated as the expected value under the probability measure $q(t)$ of the capital at risk, i.e., how much can be lost if the counterparty does not honor its part in the contract, which in our case of study corresponds to a positive payoff, discounted to the present and corrected by a loss given default (LGD) factor $1-R$:
    \begin{equation}\label{eq:CVA_exact}
        \begin{array}{ccl}
            {\rm CVA} & = & (1-R)\cdot \mathbb{E}_{q(t)}[p(t)E(t)] \\[0.1in]
             & = & (1-R)\cdot \mathbb{E}_{q(t)}\{p(t)\mathbb{E}_{P(s|t)}[v(s,t)]\} \\[0.1in]
             & = & \displaystyle (1-R)\int_0^T q(t)p(t)\int_0^\infty P(s|t)v(s,t){\rm d}s{\rm d}t.
        \end{array}
    \end{equation}
Here $R$ is the \emph{recovery rate}, defined as the fraction of the value of an asset retained after the borrower defaults.
The above CVA calculation can also be generalized to the case where the option contract has multiple underlying assets. In both the single-asset and the multi-asset cases, estimating the value of CVA within statistical error $\epsilon_S$ costs $\mathcal{O}(1/\epsilon_S^2)$.

From Equation \eqref{eq:CVA_exact} one sees that the CVA value is an integral over time and price space. Hence, a starting point for estimating the CVA is to approximate it by a sum over its value over discretized time steps $\{t_i\}_{i=0}^M$ where ${t_0=0}$ and $t_M={T}$ (note that the summation starts from $i=1$ while the definitions of $t_i$ start from $i=0$):
\begin{eqnarray}
\label{eq:cva_equation}
(1-R)\cdot\sum_{i=1}^{M} E(t_i)p(t_i)q(t_i),
\end{eqnarray}
where $p(t_i)=\exp{(-r_{t_i}t_i)}$ is the risk-free discount factor with time-dependent interest rate $r_{t_i}$ at time $t_i$, $q(t_i)$ is the probability of default between time $t_{i-1}$ and $t_{i}$. 
Moreover, $E(t_i)=\mathbb{E}_{P(s|t_i)}\left[v(s,t_i)\right]$ is the expected exposure with 
\begin{equation}
v(s,t)=\max\{s(t)-K \exp{\left(-r_t(T-t)\right)},0\},
\end{equation}
being the payoff at maturity time $T$ for a strike price $K$, assuming a single underlying asset for the European option. Here, $s(t)=s(0)\exp{\left(\sigma\xi+\left(\mu-\frac{\sigma^2}{2}\right) t\right)}$ is the asset price at time $t$, modeled as a geometric Brownian motion where $\xi\sim N(0,1)$ is a unit normal random variable, $\sigma$ is the volatility of the asset and $\mu$ the market drift, accounting for the long-term price movement trend on average at the risk-free rate. \\
To enable representation of the asset price using quantum registers, we also discretize the value of the price with $N+1$ values $\{s_j\}_{j=0}^N$. We can then approximate the distribution of the asset price fluctuation by considering the joint distribution $P(s,t)=P(s|t)P(t)$ where the marginal distribution $P(t)$ is the uniform distribution over the time period from 0 to $T$. We then define a discrete probability distribution $\mathcal{P}$ defined at each point $(s_j, t_i)$ for approximating $P(s,t)$:
\begin{equation}
\label{eq:P}
    \mathcal{P}(s_j,t_i)=\frac{1}{M\mathcal{N}}\int_{s_{j-1}}^{s_j}
    P(s|t_i){\rm d}s.
\end{equation}
where $i=1,\cdots,N$, $j=1,\cdots,M$ and $\mathcal{N}=\int_{s_0}^{s_N}P(s,t_i){\rm d}s$ is the normalization constant. 
Since the marginal distribution $P(t)$ is uniform, after discretization the marginal distribution of $\mathcal{P}$ should satisfy $\mathcal{P}(t_i)=1/M$.
This produces an approximation of the expected exposure as 
\begin{equation}
    \begin{array}{ccl}
        \tilde{E}(t_i) & = & \displaystyle \sum_{j=1}^{N}\frac{\mathcal{P}(s_j,t_i)}{\mathcal{P}(t_i)}v(s_j,t_i) \\[0.15in]
         & = & \displaystyle M\sum_{j=1}^{N}\mathcal{P}(s_j,t_i)v(s_j,t_i)
    \end{array}
\end{equation}
Combining the discretizations for both asset price and time, we arrive at the quantity to be estimated as
\begin{equation}\label{eq:CVA2}
M(1-R)\cdot\sum_{i=1}^{M}\sum_{j=1}^{N} \mathcal{P}(s_j,t_i)v(s_j,t_i)p(t_i)q(t_i).
\end{equation}

Note that in Equation \eqref{eq:CVA2} the quantities $\mathcal{P}$, $p$ and $q$ are bounded between 0 and 1, while the payoff function $v$ may not be so. Since the discretization in asset price $s$ means the value of $s$ is bounded, the value of $v$ must also be bounded. We introduce a scaling factor $C_v$ such that $v=C_v\tilde{v}$ where $\tilde{v}$ is bounded between 0 and 1. For quantities $p$ and $q$, it is possible that their values vary only subtly over their entire domains, making it hard to accurately approximate them. We therefore introduce scaling factors $C_p$ and $C_q$ such that $p=C_p\tilde{p}$ and $q=C_q\tilde{q}$. By letting $C_p>1$ and $C_q>1$ we are able to amplify the fluctuations of the functions $p$ and $q$ on their domains respectively. This leads to the final expression 
\begin{equation}\label{eq:CVA3}
\begin{array}{ccl}
\widetilde{\text{CVA}} & = & M(1-R) C_vC_pC_q\cdot \\
& & \underbrace{\sum_{i=1}^{M}\sum_{j=1}^{N} \mathcal{P}(s_j,t_i)\tilde{v}(s_j,t_i)\tilde{p}(t_i)\tilde{q}(t_i)}.
\end{array}
\end{equation}
The problem then becomes casting the bracketed term in Equation \eqref{eq:CVA3}, which is bounded between 0 and 1, as an amplitude estimation problem.

$\quad$\\
\noindent{\bf CVA instance for benchmarking.}
\label{sec:classical_benchmarks}
We consider a specific instance of the CVA problem defined on a single asset European call option: all the calculations were carried out under the specification listed in Table \ref{table:CVA_parameters}. The time points $\{t_i\}_{i=1}^M$ used for this instance are generated using the formula
\begin{equation}
\label{eq:ti}
    t_i=i\cdot\frac{T}{M},\quad i = 1, \cdots, M
\end{equation}
where $M=2^{m}$ is chosen such that the time points can be represented by the computational basis state of an $m$-qubit register. The maturity time of 6 months corresponds to $T=\frac{184}{360}\approx 0.511$, namely the number of business days (184) in the 6-month duration starting from March 5th, 2020, divided by the total number of days considered in a year under Day Count convention (360).
Using the $\emph{Actual}$ convention for day count means that all days between two dates are considered for interest accrual. If a given date is not a business day, it is adjusted according to the $\emph{Following}$ convention, which considers the first business day after the given holiday. The method used to determine the day at which payments are due is the $\emph{IMM}$ convention (International Money Market month), namely the effective dates are taken to be the third Wednesday of March, June, September and December. The $\emph{quarterly}$ frequency indicates how often payments are due. In order to compute default probabilities, we use a bootstrapping approach to recover hazard rates from market CDS quote spreads \cite{Marchioro09}, where the interest rate curve is variable over time, specifically, it is taken to be the EONIA curve.
$\quad$\\
\begin{table}[h!]
\centering
 \begin{tabular}{||c c||} 
 \hline
 Initial\ Asset\ Price & 5.0 \\
 \hline
 Strike Price & 5.5 \\ 
 \hline
 Volatility & 0.25 \\
 \hline
 Drift & 0.02 \\
 \hline
 Maturity & 6 months \\
 \hline
 Start\ Date & 05/03/2020 \\
 \hline
 CVA Recovery\ Rate & 0.415 \\
 \hline
 Notional & 1 \\ 
 \hline
 Forward\ Rate\ Curve & EONIA Curve \\
 \hline
 Hazard\ Rates & Flat\ Piecewise \\
 \hline
 CDS\ Quote\ Spreads & \begin{tabular}{c} [0.00093772, 0.00184451, \\
                0.0032286, 0.0047065, \\ 
                0.00574888, 0.00574888]\end{tabular} \\ 
 \hline
 CDS\ Tenors & [1y, 3y, 5y, 7y,  10y, 15y]\\ 
 \hline
 CDS Recovery\ Rate & 0.4125\\ 
 \hline
 CDS Settlement\ Days & 0\\ 
 \hline
 Calendar & Target \\ 
 \hline
 Day\ Count & Actual / 360 \\ 
 \hline
 Business Day\ Convention & Following \\ 
 \hline
 Date\ Generation & IMM \\ 
 \hline
 Frequency & Quarterly \\ 
 \hline
 \end{tabular}
\caption{
{Specification of the CVA instance benchmarked in this study.}}
\label{table:CVA_parameters}
\end{table}
$\quad$\\
\noindent{\emph{Benchmark value using Monte Carlo simulation.}} As a classical benchmark for numerically testing the quantum algorithm, 
    we ran Monte Carlo simulations that use $10^5$ stochastic paths mimicking the asset price fluctuation over time. The simulation results allow us to estimate the expected exposure $E(t)$ and approximate CVA by Equation \eqref{eq:cva_equation} directly, without applying the price discretization that produces $\mathcal{P}$.
The Monte Carlo simulations show that the CVA value for this particular instance is
\begin{equation}\label{eq:CVA_MC}
    {\rm CVA}_{\rm MC} = (5.599\pm 0.002)\cdot 10^{-5}. 
\end{equation}
The calculation is implemented using the Orquestra$^\text{\textregistered}$ integration with the 
\texttt{quantlib} library \footnote{https://www.quantlib.org/}. For the remainder of the paper, we will use the value of ${\rm CVA}_{\rm MC}$ as the benchmark value representing the ``exact" value of CVA, with which CVA calculations by other methods in this study are compared.

$\quad$\\
\noindent{\emph{Benchmark value with price discretization.}} The value of CVA in Equation \eqref{eq:CVA_MC} assumes discretization in time according to Equation \eqref{eq:ti} but the price can take any value from 0 to infinity. To prepare for quantum algorithm treatment, we also discretize the price, leading to the discrete distribution $\mathcal{P}$ in Equation \eqref{eq:P}.   
In Section \ref{subsec:state_prep} we provide more details on the construction of $\mathcal{P}$.

\begin{figure}
    \centering
    \includegraphics[scale=0.5]{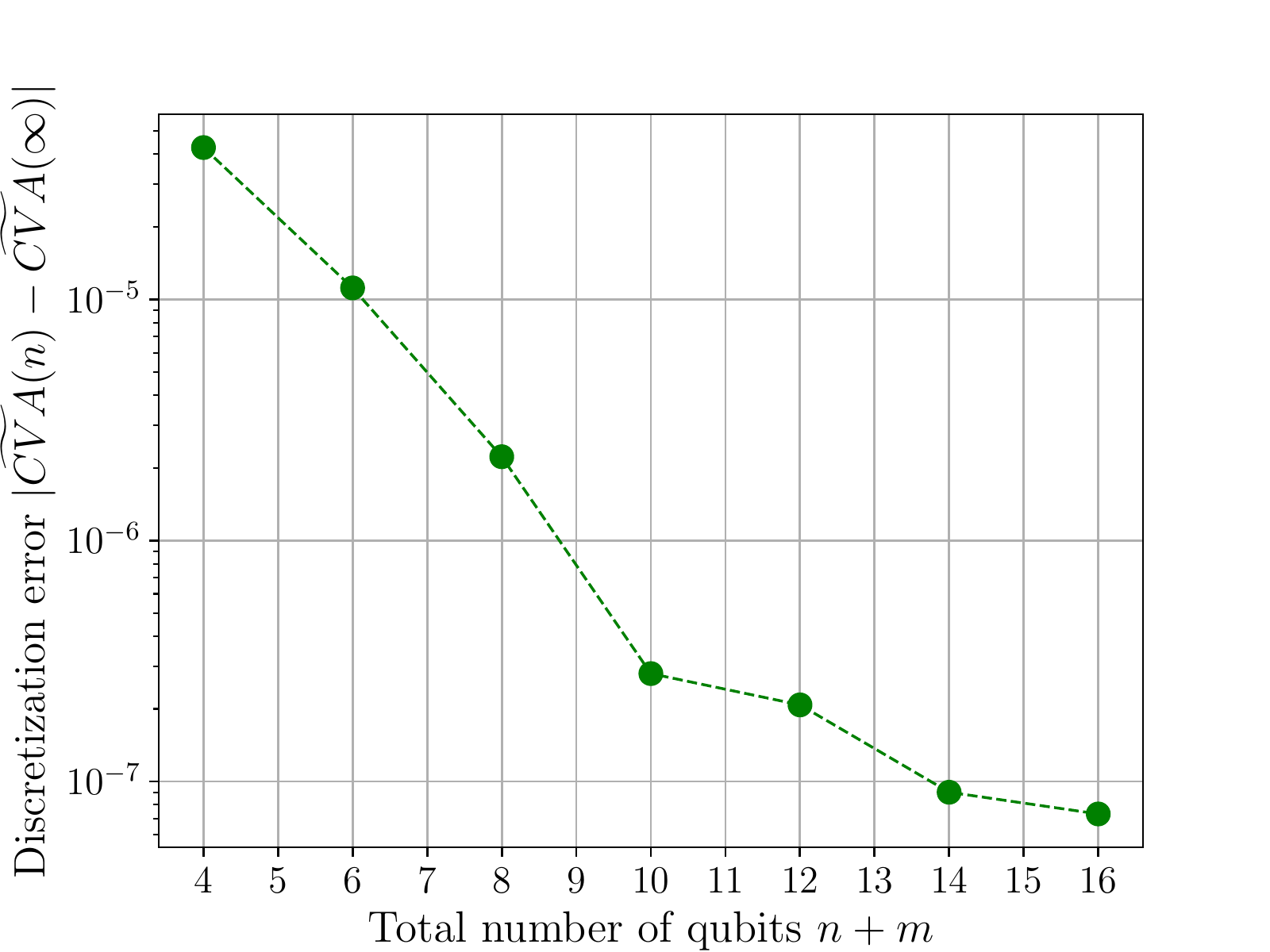}
    \caption{The convergence of $\widetilde{\rm CVA}(n)$ as $n$ grows. Here the total number of qubits is $n+m$, where the number of qubits $m$ for representing time (Equation \ref{eq:ti}) is fixed to be 2.}
    \label{fig:cva1N}
\end{figure}

Clearly, such CVA estimation based on discretized price values depends on the number $N$ of discrete price values. Let $\widetilde{\rm CVA}(n)$ be the value obtained using $N=2^n$ discrete price values evenly spaced between $s_1$ and $s_N$. Here we choose $N$ values that are powers of two for its convenience in associating with the number of qubits $n=\log_2N$ needed for encoding each price value $s_j$. Numerical results (Figure \ref{fig:cva1N}) indicate that as $n$ grows, $\widetilde{\rm CVA}(n)$ converges towards a value
\begin{equation}
\widetilde{\rm CVA}(\infty)= (5.48\pm 0.02)\cdot 10^{-5}.    
\end{equation}
The difference between $\widetilde{\rm CVA}(\infty)$ and ${\rm CVA}_{\rm MC}$ is likely due to the probability weight lost when restricting the asset price domain from $[0, \infty)$ to $[s_1,s_N]$. However, this difference accounts for only around $2\%$ relative error with respect to ${\rm CVA}_{\rm MC}$. As shown in Figure \ref{fig:cva1N}, for small values of $n$, the error due to discretization (namely the introduction of $\{(s_j,t_i)|j=1,\cdots,N;i=1,\cdots,M\}$ for representing the domain of time and price) is dominated by the number of discrete values $s_j$ that represent the price.


\section{Quantum algorithm} \label{sec:Map_to_quantum}
 
 In Section \ref{sec:cva}, the CVA calculation is broken down into 4 components.
To describe how the quantum algorithm works, we group the 4 steps into two parts: step 1 being the state preparation, steps 2-4 being the controlled rotation implementation. The final step is then assembling the operations from the previous 4 steps into a quantum circuit whose output state allows for the measurement of the bracketed term in Equation \eqref{eq:CVA3}.

$\quad$\\
\noindent{\bf State preparation.} The goal here is to (approximately) realize an operation ${G_\mathcal{P}}$
acting on $n+m$ qubits, where $n=\lceil\log_2N\rceil$ and $m=\lceil\log_2M\rceil$, such that
\begin{equation}
{G_\mathcal{P}}\ket{0^{n+m}}=\sum_{i,j}\sqrt{\mathcal{P}(s_j,t_i)}\ket{i,j}.
\end{equation}
Here $\ket{i}$ (resp.\ $\ket{j}$) is the computational basis state marking the index $i$ (resp.\ $j$) in its binary form. The setting of the marginal distribution of $\mathcal{P}$ over the asset prices is a log-normal distribution, as a consequence of choosing the geometric Brownian model as the statistical model for the underlying asset. A different distribution may be chosen for other statistical models. For example, in cases where one would like to contemplate distributions with heavy tails, Lévy distribution may be used for the marginal distribution over the asset prices. 

$\quad$\\
\noindent{\bf Controlled rotation.} The goal here is to use a quantum state $|x\rangle$ encoding an input variable $x$ as a control register for enacting a rotation on an ancilla qubits that results in a state $\sqrt{1-f(x)}|0\rangle+\sqrt{f(x)}|1\rangle$ for some function $f$. This construction is also commonly used in previous works \cite{Wocjan2009,Montanaro2015} on quantum speedups for Monte Carlo procedures. For our purpose, we introduce a controlled rotation operator for each of the steps 2-4 described in Section \ref{sec:cva}.

For representing the payoff of the option contract, we define quantum operator ${R_v}$ acting on $n+m+1$ qubits such that \begin{equation}
\label{eq:R_E}
{R_v}\ket{i}\ket{j}\ket{0}=\ket{i}\ket{j}\left(\sqrt{1-\tilde{v}(s_j,t_i)}\ket{0}+\sqrt{\tilde{v}(s_j,t_i)}\ket{1}\right).
\end{equation}  

For representing the probability of default, we introduce operator ${R_q}$ acting on $n+1$ qubits be such that
\begin{equation}
\label{eq:R_q}
{R_q}\ket{i}\ket{0}=
\ket{i}\left(\sqrt{1-\tilde{q}(t_i)}\ket{0}+\sqrt{\tilde{q}(t_i)}\ket{1}\right).
\end{equation}

For representing the discount factor, we define quantum operator ${R_p}$ acting on $n+1$ qubits such that 
\begin{equation}
\label{eq:R_p}
{R_p}\ket{i}\ket{0}=\ket{i}\left(\sqrt{1-\tilde{p}(t_i)}\ket{0}+\sqrt{\tilde{p}(t_i)}\ket{1}\right).
\end{equation}

\begin{figure*}
    \centering
    \includegraphics[scale=0.4]{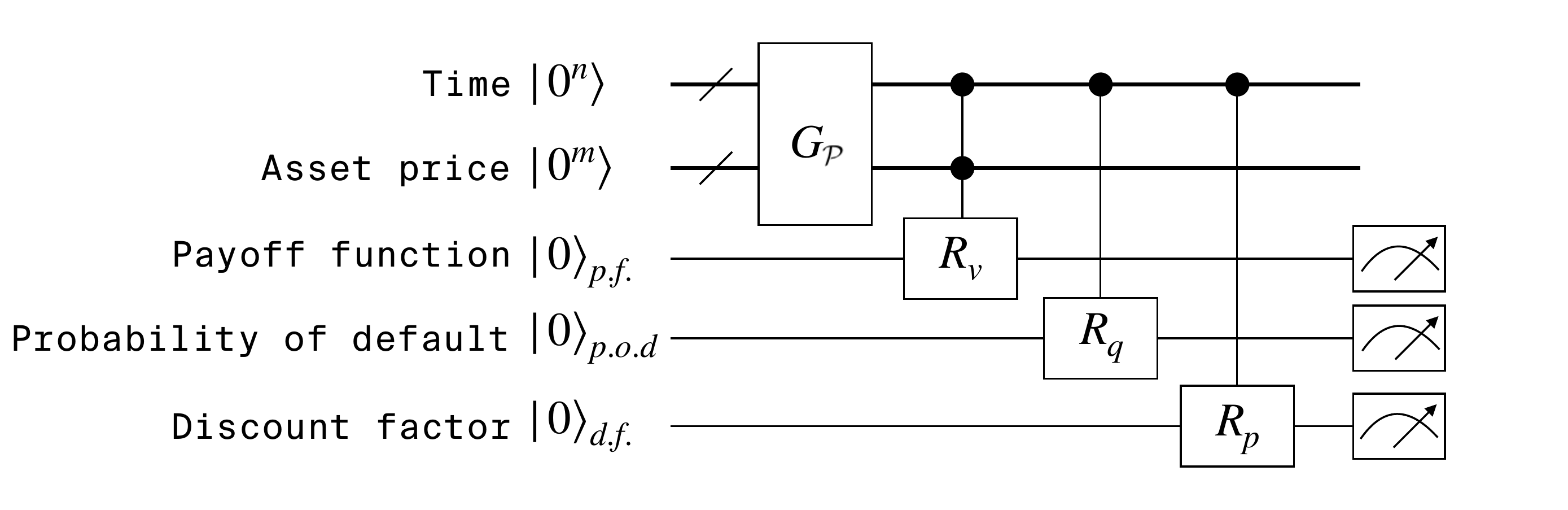}
    \caption{The proposed (ideal) quantum circuit to perform credit value adjustment as described in Section \ref{sec:Map_to_quantum}.}
    \label{fig:cva_circuit}
\end{figure*}

$\quad$\\
\noindent{\bf Quantum circuit assembly.} We could then describe a procedure for estimating the bracketed quantity in \eqref{eq:CVA3} as the following (Figure \ref{fig:cva_circuit}): 
\begin{enumerate}
\item Start with two quantum registers, one of $n$ qubits and the other of $m$ qubits and generate the quantum state
\begin{equation}
\sum_{i=1}^{M}\sum_{j=1}^{N}\sqrt{\mathcal{P}(s_j,t_i)}\ket{i}\ket{j}.
\end{equation}
using the operator ${G_\mathcal{P}}$.


\item Add an ancilla qubit in $\ket{0}$ for storing the payoff function ($p.f.$) and apply the operator ${R_v}$ to produce an entangled state 
\begin{eqnarray}
&&\sum_{i=1}^{M}\sum_{j=1}^{N}\sqrt{\mathcal{P}(s_j,t_i)}\ket{i}\ket{j}\otimes\nonumber \\
&&\displaystyle
\left(\sqrt{1-\tilde{v}(s_j,t_i})\ket{0}_{p.f.}+\sqrt{\tilde{v}(s_j,t_i)}\ket{1}_{p.f.}\right). 
\end{eqnarray}

\item Add another ancilla qubit in $\ket{0}$ for storing the probability of default ($p.o.d.$) and apply the operator ${R_q}$ onto the first register and the new ancilla qubit to produce the state 
\begin{eqnarray}
&&\sum_{i=1}^{M}\sum_{j=1}^{N}\sqrt{\mathcal{P}(s_j,t_i)}\ket{i}\ket{j}\otimes\nonumber\\
&&\left(\sqrt{1-\tilde{v}(s_j,t_i})\ket{0}_{p.f.}+\sqrt{\tilde{v}(s_j,t_i)}\ket{1}_{p.f.}\right)\otimes\nonumber\\
&&\left(\sqrt{1-\tilde{q}(t_i)}\ket{0}_{p.o.d.}+\sqrt{\tilde{q}(t_i)}\ket{1}_{p.o.d.}\right).\nonumber\\
\end{eqnarray}

\item Add another ancilla qubit in $\ket{0}$ for storing the discount factor ($d.f.$) and apply the operator $R_p$ onto the first register and the new ancilla qubit to produce the final state
\begin{eqnarray}
\ket{\xi}&=&\sum_{i=1}^{M}\sum_{j=1}^{N}\sqrt{\mathcal{P}(s_j,t_i)}\ket{i}\ket{j}\otimes\nonumber\\
&&\left(\sqrt{1-\tilde{v}(s_j,t_i})\ket{0}_{p.f.}+\sqrt{\tilde{v}(s_j,t_i)}\ket{1}_{p.f.}\right)\otimes\nonumber\\
&&\left(\sqrt{1-\tilde{q}(t_i)}\ket{0}_{p.o.d}+\sqrt{\tilde{q}(t_i)}\ket{1}_{p.o.d}\right)\otimes\nonumber\\
&&\left(\sqrt{1-\tilde{p}(t_i)}\ket{0}_{d.f.}+\sqrt{\tilde{p}(t_i)}\ket{1}_{d.f.}\right)
\label{eq:ansatz_state}
\end{eqnarray}

\item Let $\Pi$ be projector onto the subspace where the $d.f.$, $p.f.$ and $p.o.d.$ ancilla qubits are all in the state $\ket{1}$. More explicitly, we have  
\begin{eqnarray}
\Pi&=&\ket{1}\bra{1}_{d.f.}\otimes\ket{1}\bra{1}_{p.f.}\otimes\ket{1}\bra{1}_{p.o.d.} \nonumber \\
&=&\frac{1}{8}(I-Z_{d.f.}-Z_{p.f.}-Z_{p.o.d.} \nonumber \\
&&+Z_{d.f.}Z_{p.f.}+Z_{d.f.}Z_{p.o.d.}+Z_{p.f.}Z_{p.o.d.} \nonumber \\
&&-Z_{d.f.}Z_{p.f.}Z_{p.o.d.}), 
\label{eq:projector}
\end{eqnarray}
which is a linear combination of Pauli operators that can be measured directly and simultaneously on the quantum processor.
\end{enumerate}
We observe that the quantity desired in Equation \eqref{eq:CVA3} can be obtained by 
\begin{equation}
\sum_{i=1}^M\sum_{j=1}^N \mathcal{P}(s_j,t_i)\tilde{v}(s_j,t_i)\tilde{p}(t_i)\tilde{q}(t_i)=\bra{\xi}\Pi\ket{\xi},
\label{eq:exp_val_problem}
\end{equation}
which can be estimated within a sampling error of $\epsilon_S$ in time $\mathcal{O}(1/\epsilon_S)$ using amplitude estimation. \\

The rest of the paper is organized as follows.
In Section \ref{sec:quantum_circuit} we describe in details the quantum circuit construction used for preparing the state $\ket{\xi}$, using a particular circuit as an example. The quantum circuit construction involves the following operations: 
\begin{itemize}
\item Operator $G_\mathcal{P}$ for state preparation (Section \ref{subsec:state_prep});
\item Controlled operations $R_v$, $R_q$, $R_p$  (Section \ref{sec:crca_section}).
\end{itemize}

The proposed quantum circuit for the algorithm can be found in Figure \ref{fig:cva_circuit}. In Appendix \ref{sec:ELF} we proceed to describe how the quantum circuit design is used for amplitude estimation by engineered likelihood function (ELF) \cite{2006.09350}.

\section{Quantum circuit}
\label{sec:quantum_circuit}
 
Existing state preparation techniques \cite{quant-ph/0208112,10.1145/780542.780546,0801.0342,10.1145/2493252.2493256} rely on the ability to perform operations such as evaluating arithmetic expressions \cite{0801.0342}, computing the integral of a probability density function over an interval \cite{quant-ph/0208112}, and extracting elements of a sparse matrix \cite{10.1145/780542.780546}. These operations are either assumed to be supplied through oracles, in which case their implementations in terms of elementary quantum operations on a quantum computer are entirely not taken in to account, or assumed to be realizable efficiently via well-known techniques such as reversible logic synthesis (RLS) \cite{Vedral1996,Beckman1996,10.5555/2012086.2012090,quant-ph/0410184,d.m.millerd.maslovg.w.dueck2003, soekenm.dueckg.w.millerd.m.2016}. However, although RLS is efficient in an asymptotic sense if the underlying function can be realized by a polynomial-time classical procedure, concrete resource estimations show that it is costly compared with other parts of quantum algorithms, motivating alternative approaches \cite{10.5555/3179320.3179329}. In this paper we also perform numerical experiments to illustrate how costly RLS can be with even small examples of CVA calculations (Section \ref{sec:crca_section}).

Instead of RLS, which is capable of enabling both the controlled rotation and the state preparation \footnote{With RLS, consider the state preparation process $|i\rangle|0\rangle\rightarrow|i\rangle|p(i)\rangle\rightarrow|i\rangle|p(i)\rangle(\sqrt{1-p(i)}|0\rangle+\sqrt{p(i)}|1\rangle)$ and post selection on the state $|1\rangle$ for the last qubit. With proper uncomputation, one can then generate the state $\Sigma_i\sqrt{p(i)}|i\rangle$ which is relevant in the context of this paper.}, in this work we identify opportunities for realizing both with circuits of much shorter depths. For state preparation, we investigate two alternatives: quantum circuit Born machine \cite{zhu2019training} and quantum circuit construction based on matrix product states \cite{LeiWang18}. We show that highly accurate approximations are possible with circuits that are much shallower than those produced from RLS. We note that there are also other state preparation schemes \cite{DallaireDemers2018,Zoufal2019,Romero2020} inspired by generative adversarial networks that can also yield more efficient state preparation protocols than RLS.

For the purpose of illustration, we consider a specific instance of CVA estimation using Equation \eqref{eq:CVA3} with parameters listed in Table \ref{table:circuit_params}.

The classical benchmark value (Section \ref{sec:cva}) for this instance is 
\begin{equation}
\label{eq:CVA_2}
\widetilde{\rm CVA}(2)=1.223\cdot 10^{-5}.
\end{equation}
The large discrepancy between $\widetilde{\rm CVA}(2)$ and ${\rm CVA}_{\rm MC}$ in Equation \eqref{eq:CVA_MC} can be mostly attributed to discretization error due to finite $n$ (Figure \ref{fig:cva1N}). In Section \ref{sec:quantum_cva} we will compute the value $\widetilde{\rm CVA}_Q$ produced by the quantum circuit for this instance, compare it with $\widetilde{\rm CVA}(2)$ and discuss the sources of error.
$\quad$\\
\begin{table}[h!]
\centering
 \begin{tabular}{||c | c||} 
 \hline
 Number of qubits $m$ for encoding time & $2$ \\
 Number of qubits $n$ for encoding price & $2$ \\
 Scaling constant for payoff $C_v$ & $1.8201814$ \\
 Scaling constant for default probability $C_q$ & $0.0002038$ \\
 Scaling constant for discount factor $C_p$ & $1$ \\
 \hline
 \end{tabular}
 \caption{Parameters of the example quantum circuit considered in Section \ref{sec:quantum_circuit}. The other parameters related to the CVA instance are shown in Table \ref{table:CVA_parameters}.}
 \label{table:circuit_params}
 \end{table}

\subsection{State preparation}
\label{subsec:state_prep}
\begin{figure*}
    \centering
    \includegraphics[scale=0.6]{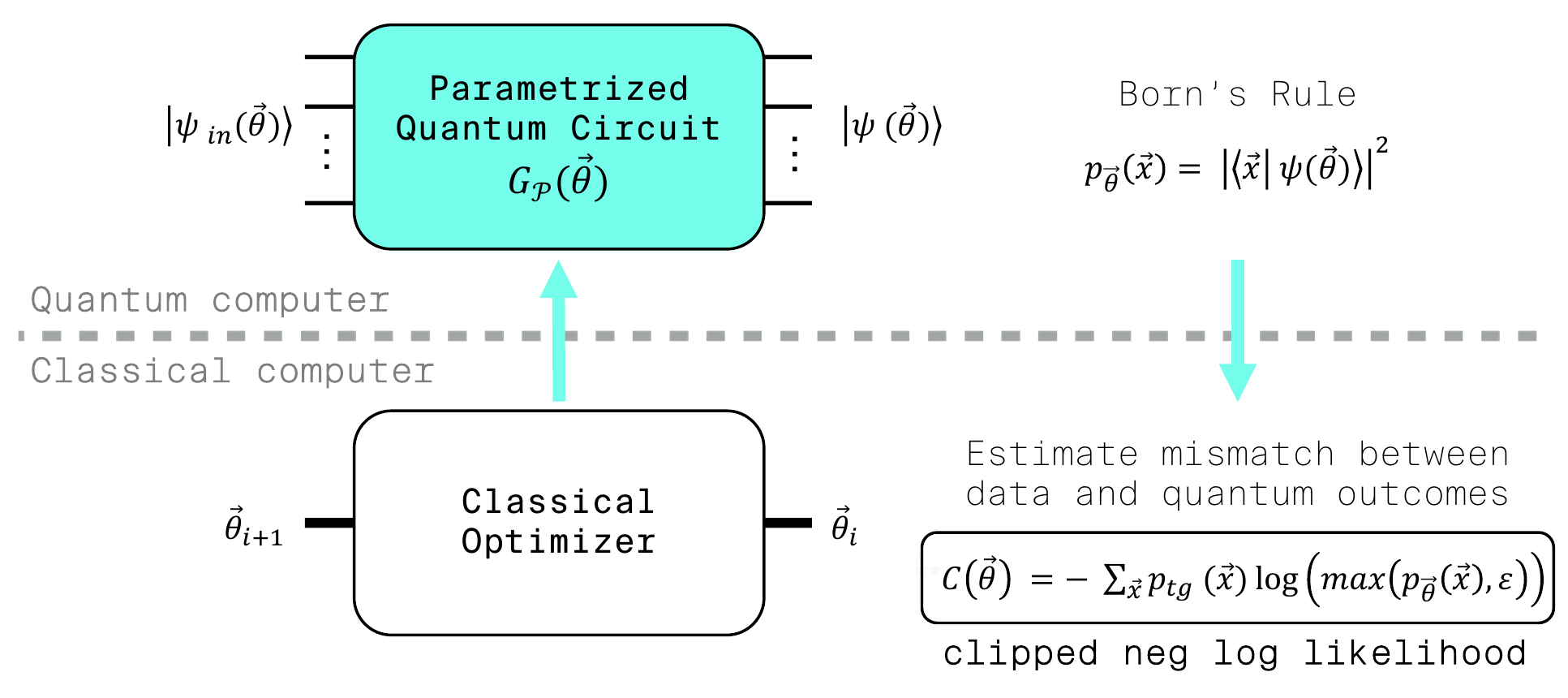}
    \caption{Quantum circuit Born machine for learning the state preparation circuit $G_\mathcal{P}$.}
    \label{fig:qcbm}
\end{figure*}
In the context of the CVA problem (Section \ref{sec:Map_to_quantum}), the role of state preparation is to implement the $G_\mathcal{P}$ operator, thus preparing a quantum state loading into it the target distribution $p_{tg}({x})=\mathcal{P}(s_j,t_i)$ which is the discretized joint distribution of time and asset price. The circuit acts on two registers of qubits encoding time and asset price respectively, where the qubit states are represented as bitstrings whose first part refers to the time register and the following to the price register. 
Given that we are considering asset fluctuations modeled by geometric Brownian motion, at each point in time such distribution is the log-normal distribution, namely:

\begin{equation}
    P(s|t) = \dfrac{1}{s\sqrt{2\pi\sigma^2}}\exp\bigg[-\bigg(\dfrac{\ln{s} - \ln{s_0} - (\mu - \frac{\sigma^2}{2})t}{\sqrt{2\sigma^2t}}\bigg)^2\bigg],
\label{eq:lognormal_distribution}
\end{equation}
where $\sigma$ is the price volatility, $\mu$ represents the market drift which accounts for the long term price movement trend, and $s_0$ is the initial asset price. 

The target distribution of $\mathcal{P}$ in Equation \eqref{eq:P} is obtained via classical Monte Carlo simulation of the asset prices and then discretized to the available quantum states $|{x}\rangle = |i,j\rangle$. Specifically, $10^5$ trajectories of asset prices dynamics from time 0 to maturity $T$ are computed, simulating geometric Brownian motion on a fine time grid. Then, the asset price distributions at the time steps $t_i$ defined by Equation \eqref{eq:ti} are extracted from the above simulation: for each time step $t_i$ we obtain a log-normal peak as in Equation \eqref{eq:lognormal_distribution}. 
We choose the range of the discrete $s_j$ values such that the smallest value $s_1 = \max\{ \hat{\mu}-3\hat{\sigma},0\}$ and the largest value $s_N=\hat{\mu}+3\hat{\sigma}$, where $\hat{\mu}$ and $\hat{\sigma}$ are the sample mean and sample standard deviation of the price values produced by the Monte Carlo simulations. We then calculate $\mathcal{P}$ by binning the price data from the Monte Carlo simulation yielding ${\rm CVA}_{\rm MC}$ according to the different $s_j$ values, which enabled the estimation of CVA according to Equation \eqref{eq:CVA3}.

\subsubsection{Quantum Circuit Born Machine}
\label{subsec:QCBM}
Quantum Circuit Born Machine (QCBM) \cite{zhu2019training} has been shown to learn and load a target probability distribution $p_{tg}$ into a quantum state. The structure of this hybrid quantum-classical algorithm is depicted in Figure \ref{fig:qcbm}. 
The subroutine running on the quantum computer consists of training a parametrized quantum circuit, depending on some parameters $\vec{\theta}$ and encoding a probability distribution $p_{\vec{\theta}}$. Indeed, the output state of the circuit $\ketsmall{\psi(\vec{\theta})}$ contains in its amplitudes the probability distribution, according to Born's rule: 
\begin{equation}
    \label{born_rule}
    p_{\vec{\theta}}({x}) = |\braket{{x}}{\psi(\vec{\theta})}|^2,
\end{equation}
where $|{x}\rangle=|i,j\rangle$ represents a computational basis state that encodes a point on the discretized domain of the target probability distribution $p_{tg}$.
The circuit parameters $\vec{\theta}$ are tuned in order to find the optimal set $\vec{\theta}^*$ such that $p_{\vec{\theta}}$ is as close as possible to $p_{tg}$.
The process of learning the parameters is carried out by a classical optimizer: its goal is to minimize a cost function which quantifies the difference between the two probability distribution into play. We choose to employ an evolutionary and derivative free strategy as the classical optimizer, namely the Covariance Matrix Adaptation Evolution Strategy (CMA-ES \cite{hansen2006cma}). As for the cost function, we use the clipped negative log-likelihood \cite{zhu2019training}, defined as follows: 
\begin{equation}
    C(\vec{\theta}) = -\sum_{{x}} p_{tg}({x}) \log(\max(p_{\vec{\theta}}({x}),\epsilon), 
    \label{eq:log-likelihood}
\end{equation}
where $\epsilon$ is a small parameter to avoid singularity.
There are several options available for both the optimizer and the cost function, whose efficiency is highly dependent on the specific problem instance. 

\begin{figure}[H]
  \includegraphics[width=\linewidth]{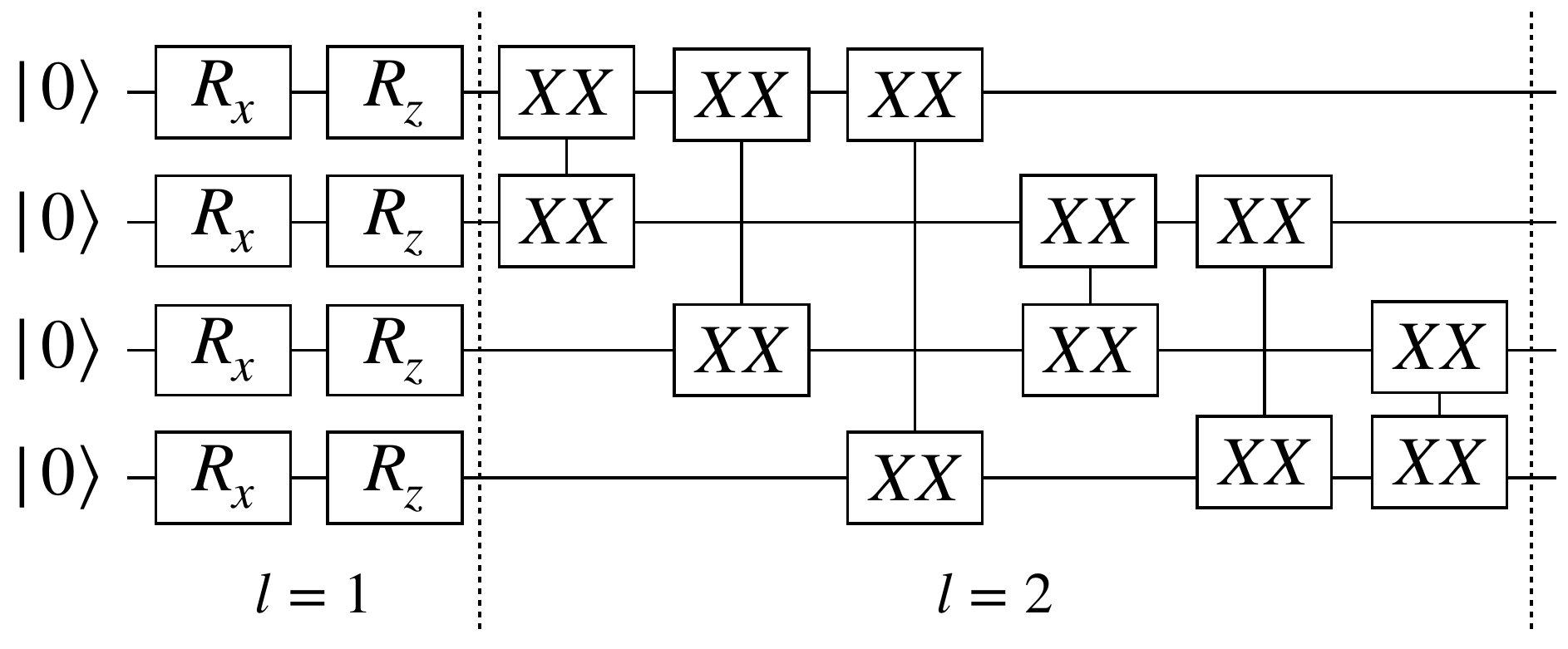}
  \caption{{The ansatz circuit for realizing the operator $G_\mathcal{P}$, with 14 variational parameters (one for each gate). It is composed of a single qubit layers ($l=1$) and an entangling layer ($l=2$), so that overall it has $n_{\rm layers}=2$.}}
  \label{fig:qcbm-ansatz}
\end{figure}

The QCBM is able to learn the desired distribution $\mathcal{P}(s_j,t_i)$, by undergoing a training process that tunes the parameters of the quantum circuit so that the desired target is loaded into a quantum state. The ansatz we consider for the parametrized quantum circuit, defined on 4 qubits, is shown in Figure \ref{fig:qcbm-ansatz}. \\

\begin{figure*}
    \centering
    \includegraphics[scale=0.35]{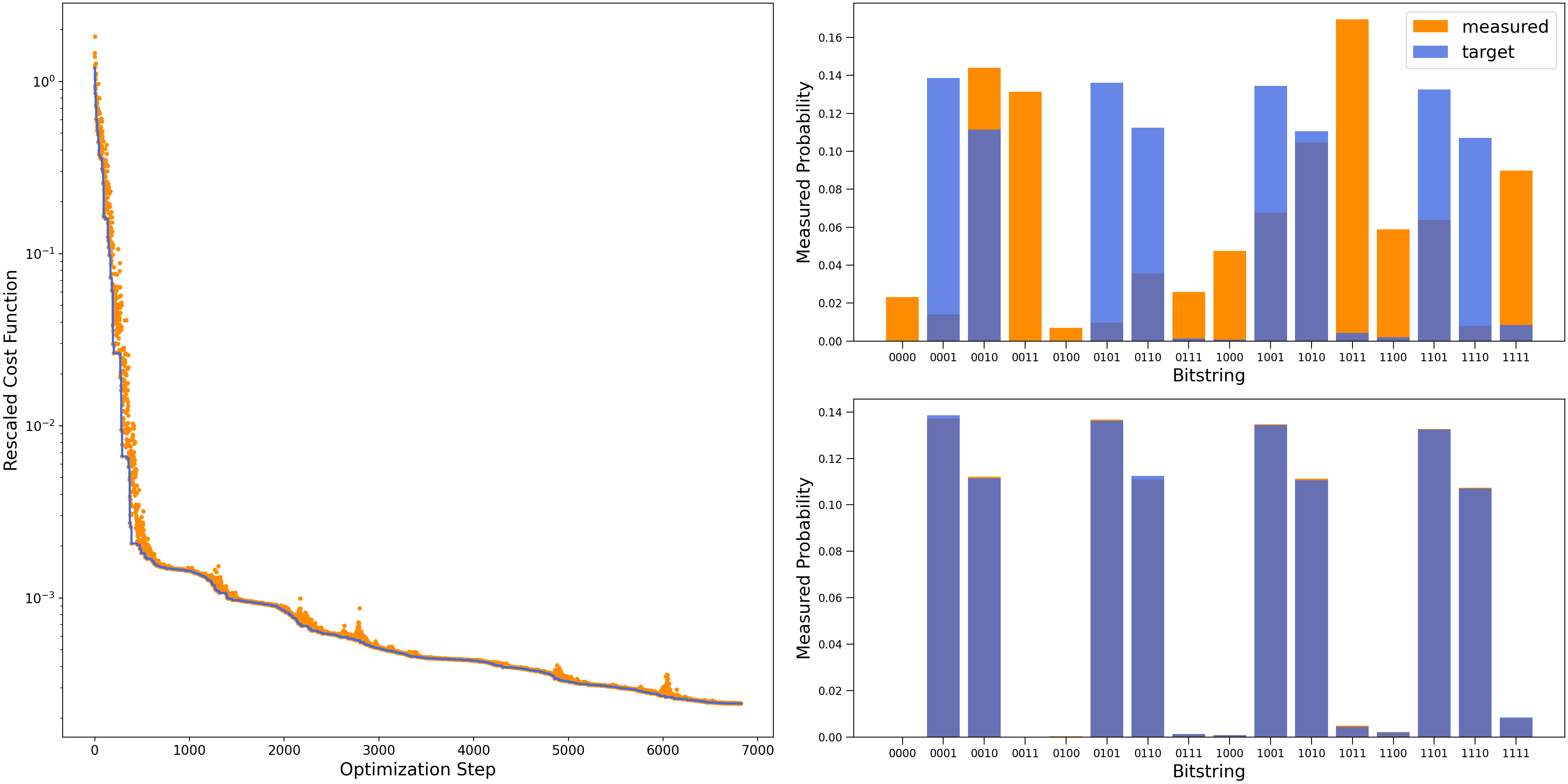}
    \caption{{The training process of the parametrized quantum circuit in the single asset case, where the ansatz is defined on 4 qubits and has 14 gates. On the left, the plot shows the convergence of the (rescaled) cost function towards its ideal value. The orange dots mark the cost function value at each optimization step, whereas the blue line connects only points with progressively decreasing values, thus highlighting the minimization trend.  On the right, histograms display the probability distributions encoded in the quantum state before (top) and after (bottom) the training.}}
    \label{fig:qcbm-training}
\end{figure*}
\quad\\
It is composed of two layers, where the first layer ($l=1$) is made of single qubit rotations and the second one ($l=2$) uses two qubit gates to introduce entanglement. The depth of the circuit can be varied as needed, so that each layer of one(two)-qubit(s) gates is identified by an odd(even) index $l$, with $l=1,\cdots,n_{\rm layers}$. The types of one-qubit gates used in the QCBM circuit are chosen to be $X$ and $Z$ rotations, whereas for the two-qubits gates we use the $XX$ coupling gate, which is implemented natively in ion-trap quantum computers. Specifically, the gates are defined as follows:
 \begin{equation}
 \begin{aligned}
    & R_{x,z}(\theta) = \exp\big(-i \tfrac{\theta}{2} \sigma_{x,z}\big)\\
    & XX(\theta) = \exp\big(-i\theta\sigma_x \otimes \sigma_x\big).
\end{aligned}
\end{equation}
The qubit connectivity is assumed to be all-to-all, so that each pair of qubits undergoes an $XX$ transformation. The depth of the circuit grows according to the circuit size: the larger the number of qubits, the deeper the circuit needs to be in order to be trained successfully (see Appendix \ref{appendix:training_qcbm}). For the 4-qubit instance in Figure \ref{fig:qcbm-ansatz}, one layer of tunable entangling gates is enough for the QCBM to be able to learn the target distribution. 
Figure \ref{fig:qcbm-training} shows a typical training curve of the cost function, as the iterations proceed. Note that the cost function has been rescaled so that its expected value for a successful training is zero.


\begin{figure*}
    \centering
    \includegraphics[scale=0.36]{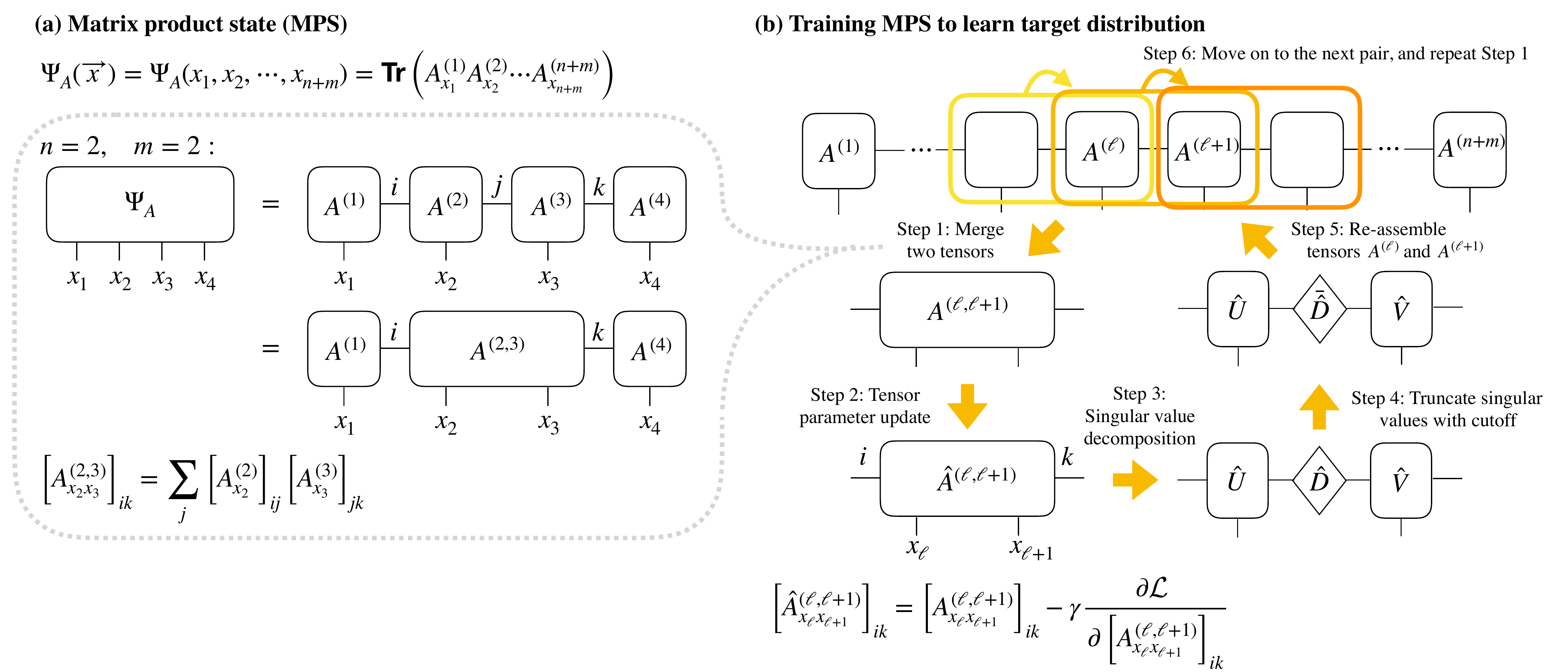}
    \caption{Matrix product state (MPS) and its usage for learning a target distribution. \textbf{(a)} Definition of an MPS $\Psi_A({x})$ and process for contracting two tensors in the MPS. \textbf{(b)} Process for training MPS to learn a target distribution. Note that in Step 5 one could either ``go left" by letting $A^{(\ell)}=\hat{U}\cdot\bar{\hat{D}}$, $A^{(\ell+1)}=\hat{V}$ or ``go right" by letting $A^{(\ell)}=\hat{U}$ and $A^{(\ell+1)}=\bar{\hat{D}}\cdot\hat{V}$. Details for evaluating the gradient of the cost function $\mathcal{L}$ are shown in \cite{LeiWang18}.}
    \label{fig:mps}
\end{figure*}

\subsubsection{Matrix Product State}\label{subsec:MPS}
An alternative method to using QCBM is to leverage Matrix Product State (MPS) \cite{LeiWang18}, which is a tensor network that has been used for mimicking correlations between different variables. An MPS is represented by a sequence of tensors $A^{(1)}$, $A^{(2)}$, $\cdots$, $A^{(n+m)}$. Each tensor $A^{(\ell)}$ depends on the value of $x_{\ell}$, which is the $\ell$-th element of ${x} = |i,j\rangle$. When evaluated at a particular value of $x_\ell$, $A^{(\ell)}$ becomes a matrix (except for when $\ell=1$ or $\ell=n+m$, where it becomes a vector), denoted as $A^{(\ell)}_{x_\ell}$. The MPS can then be defined as (Figure \ref{fig:mps}a)
\begin{equation}
\label{eq:mps_Psi}
    \Psi_A({x})=\text{Tr}\left(A^{(1)}_{x_1}A^{(2)}_{x_2}\cdots A^{(n+m)}_{x_{n+m}}\right).
\end{equation}
The equation above assumes that the $A_{x_{\ell}}^{(\ell)}$ objects have compatible dimensions so that they can be multiplied together properly. If $A^{(\ell)}_{x_{\ell}}$ is a matrix of dimension $d_{\ell}\times d_{\ell+1}$ and $A^{(\ell+1)}_{x_{\ell+1}}$ is a matrix of dimension $d_{\ell+1}\times d_{\ell+2}$, then $d_{\ell+1}$ is the \emph{bond dimension} between $A^{(\ell)}_{x_{\ell}}$ and $A^{(\ell+1)}_{x_{\ell+1}}$.

An MPS can represent a quantum wavefunction $\Psi_A({x})$ that approximates a target distribution $p_{tg}$ in the same way as QCBM, namely $p_A({x})=|\Psi_A({x})|^2  \approx p_{tg}({x})$. A concrete way to measure the closeness between the MPS output $p_A$ and the target distribution $p_{tg}$ is by computing a negative log-likelihood function over a set $S$ of samples ${x}$ generated from $p_{tg}$:
\begin{equation}
\label{log_like}
\mathcal{L}(A) = -\frac{1}{|S|}\sum\limits_{{x}\in {S}} \log p_A({x}).
\end{equation}
The likelihood function is similar to the cost function in Equation \ref{eq:log-likelihood} in the sense that averaging over $S$ approximates the expectation over the target distribution $-\sum_{{x}}p_{tg}({x})\log p_A({x})$.

\begin{figure}
  \makebox[-0.3in]{}
  \includegraphics[scale=0.47]{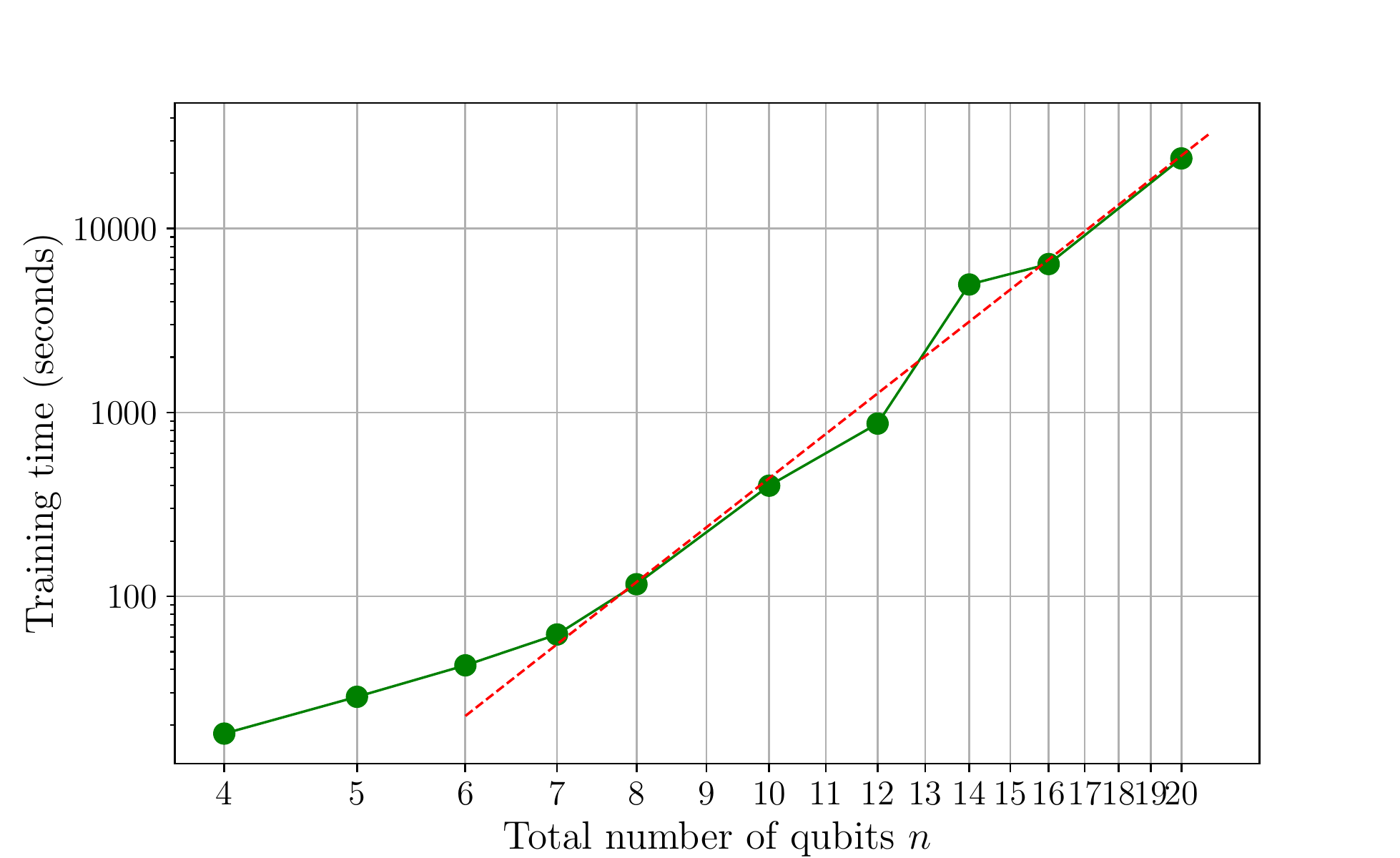}
  \caption{Runtime for training the MPS versus the number of qubits. The scales for both axes are logarithmic. The red dashed line is a linear regression with correlation coefficient being 0.987 and the slope being roughly 5.827.}
  \label{fig:RunTime_Nqubits_No20}
\end{figure}

In order to minimize the negative log-likelihood function, we adopt an iterative scheme. A summary of the scheme is provided in Figure \ref{fig:mps}b. This method builds on the connection between unsupervised generative modeling and quantum physics, where MPS is employed as a model to learn the probability distribution of a given data set with an algorithm which resembles {Density\ Matrix\ Renormalization\ Group} (DMRG), which is an efficient algorithm that attempts to find the MPS wavefunction corresponding to the ground state for a given Hamiltonian. Relying on the ability to efficiently evaluate gradients of the objective function \cite{LeiWang18}, we can iteratively improve the MPS approximation of the target distribution. The iterations proceed until one of the three scenarios: 1) a threshold for the KL divergence between the model output and the training data is reached, 2) the difference in the objective function between two adjacent training steps is below a threshold, or 3) a maximum number of iterations is reached. In Figure \ref{fig:RunTime_Nqubits_No20} we show results for training the MPS up to $n_{\rm qubits}=n+m=20$ qubits.
From the data, the empirical scaling of the MPS training cost is roughly $O(n_{\rm qubits}^6)$.

\begin{figure*}
    \centering
      \includegraphics[scale=0.4]{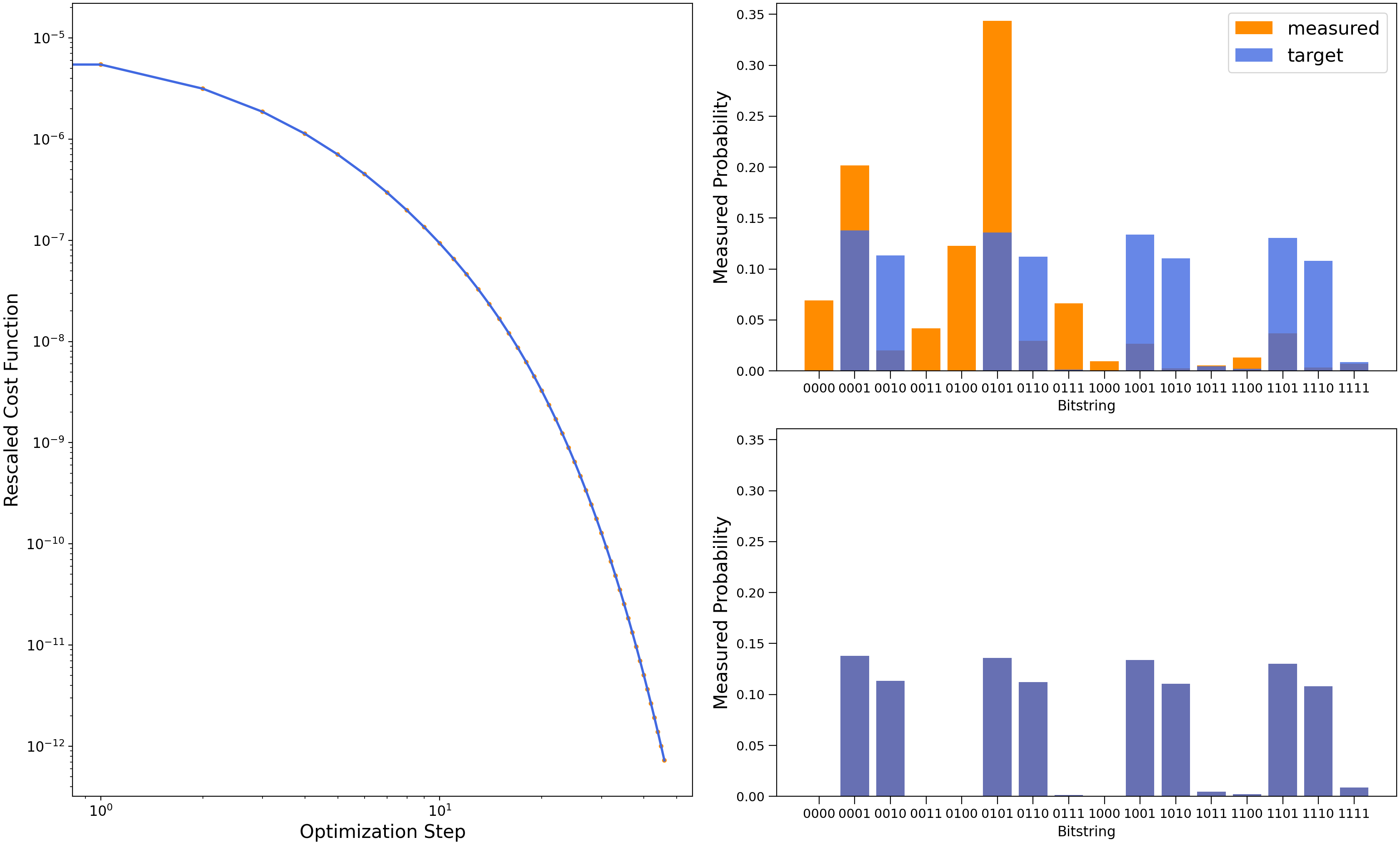}
      \caption{{The training process of the MPS model in the single asset case, where the ansatz is defined on 4 qubits and has 14 gates. On the left, the plot shows the convergence of the cost function towards its ideal value with y-axis rescaled logarithmically. The dots mark the cost function value at each optimization step, whereas the line connects only points with progressively decreasing values, thus highlighting the minimization trend.  On the right, histograms display the probability distributions encoded in the quantum state after the first training step (top) and the one after the final step of training (bottom).}}
      \label{fig:MPS_training}
\end{figure*}

\begin{table}[h!]
\centering
 \begin{tabular}{||c c c c||} 
 \hline
 $n_{\rm qubits}$ & $t_{\rm train}$ & KL & $n_{\rm iters}$ \\ [0.5ex] 
 \toprule
 4 & 17.94 & $3.21\cdot 10^{-6}$ & 2 \\ 
 \hline
 5 & 28.44 & $4.11\cdot 10^{-6}$ & 2 \\
 \hline
 6 & 42.17 & $4.53\cdot 10^{-6}$ & 2  \\
 \hline
 7 & 62.06 & $1.62\cdot 10^{-6}$ & 3  \\
 \hline
 8 & 116.46 & $8.60\cdot 10^{-5}$ & 3  \\
 \hline
 10 & 399.99 & $2.07\cdot 10^{-2}$ & 5  \\
 \hline
 12 & 870.41 & $1.5\cdot 10^{-2}$ & 7  \\
 \hline
 14 & 4967.78 & $7.04\cdot 10^{-2}$ & 19  \\
 \hline
 16 & 6417.41 & 0.86 & 22  \\
 \hline
 20 & 24084.47 & 3.20 & 25 \\ [1ex] 
 \hline
\end{tabular}
\caption{{Results for MPS training. Here $n_{\rm qubits}=n+m$ is the number of qubits encoding the joint time and price values, $t_{\rm train}$ is the total time for training the MPS (in seconds). The maximum bond dimension of the MPS in all cases is limited to 2. The values on this table has been obtained with a threshold for KL divergence of $3\cdot 10^{-5}$.}}
\label{table:Runs_results2}
\end{table}


In order to make the MPS construction useful for the CVA calculation, as well as to compare QCBM and MPS in a fair way, we need an explicit recipe for generating quantum circuits that prepare the MPS.
The basic idea is to use a sequence of Singular Value Decomposition (SVD) steps to transform the MPS into an orthogonal form in which each tensor is an isometry and hence can be embedded into a unitary operator. Then we decompose these unitary operators into CNOT and single-qubit gates by  the method in \cite{circuit_synthesis}. When the MPS has $n$ sites and bond dimension $D$, there are at most $n-\lceil \log_2(D) \rceil$ such unitary operators, where each of them acts on at most $\lceil \log_2(D) \rceil + 1$ qubits and can be implemented with $O(D^2)$ two-qubit gates. As a consequence, the quantum circuit for preparing the MPS contains $O(nD^2)$ two-qubit gates. Note that it remains unknown how many CNOT and single-qubit gates are needed exactly to implement a general $k$-qubit unitary operator for $k \ge 3$, and the best known results are lower and upper bounds on them. So we can only give lower and upper bounds on the numbers of elementary gates in the circuit for preparing the MPS (except for the case $D=2$) in Table \ref{table:MPS_depth_comparison}. See Appendix \ref{appendix:qc_mps} for more details about encoding MPS into quantum circuits.

We are now able to compare the two methods proposed so far for the state preparation task. Specifically, we focus on the aforementioned 4-qubits instance of the QCBM with different depths, i.e. $n_{\rm layers}=[2,4,6,8,10]$, which corresponds to a 4-site MPS with the same values of bond dimension $D$. We also consider a 6-qubits QCBM instance and its analogous MPS, varying $n_{\rm layers}$ and $D$ as in the former setting. The target distribution corresponds to Equation \eqref{eq:P} with $P(s|t)$ as in Equation \eqref{eq:lognormal_distribution}.
Tables \ref{table:QCBM_depth_comparison} and \ref{table:MPS_depth_comparison} show the results of the comparison between QCBM and MPS-based circuits, where we report the number of one- and two- qubit gates (in terms of CNOTs for a fair comparison), the number of iterations until convergence of the cost function minimization and the corresponding values of KL divergence, i.e. training accuracy. From this data, we can draw the following conclusions, that hold for both cases under examination:
\begin{itemize}
    \item Given a circuit depth and a target accuracy, MPS is able to reach the desired KL divergence value with very few iterations, while the QCBM requires many more optimization steps.
    \item Given a circuit depth, we observe that in some cases the QCBM is able to reach a lower KL divergence than MPS, provided we use a sufficiently high number of iterations. Additional numerics suggests that the QCBM is able to beat the training accuracy of MPS, which instead reaches a plateau that prevents the KL value to decrease further when the number of the optimization steps is increased. 
    \item Given a fixed number of qubits, we compare MPS and QCBM depths in terms of CNOTs. With small $n_{\rm layers}$ or $D$, we see that MPS is shallower than the QCBM, but as $n_{\rm layers}$ or $D$ are increased there is an inversion point - whose exact location depends on the circuit size - where QCBM becomes shallower than MPS.
\end{itemize}
We note that some of the recent works on using MPS for describing continuous probability distributions \cite{holmes2020efficient} can significantly improve the training cost of MPS or avoid training altogether. This will also affect the comparison with QCBM.
\begin{table}[H]
\centering
\begin{tabular}{||c| c c c c c||} 
 \hline
 \textbf{MPS}& $D$ \hspace{5mm}& CNOT \hspace{5mm}&  1-qb \hspace{5mm}& $n_{\rm iters}$ \hspace{5mm}& KL \\ [0.5ex]
 \toprule
 \multirow{6.5}{*}{\begin{sideways}$n_{\rm qubits}=4$\end{sideways}} 
 & 2 & 9 & 19 & \begin{tabular}{c} 2 \end{tabular} & \begin{tabular}{c} $1.15\cdot 10^{-4}$ \end{tabular} \\\cline{2-6}
  & 4 & [28, 40] & [42, 80] & \begin{tabular}{c} 2 \end{tabular} & \begin{tabular}{c} $3.21\cdot 10^{-6}$ \end{tabular} \\\cline{2-6}
  & 6 & [61, 100] & [85, 184] & \begin{tabular}{c} 2 \end{tabular} & \begin{tabular}{c} $3.21\cdot 10^{-6}$ \end{tabular} \\\cline{2-6}
  & 8 & [61, 100] & [85, 184] & \begin{tabular}{c} 2 \end{tabular} & \begin{tabular}{c} $3.21\cdot 10^{-6}$ \end{tabular} \\\cline{2-6}
  & 10 & - & - & \begin{tabular}{c} - \end{tabular} & \begin{tabular}{c} - \end{tabular} \\
  \toprule
  \multirow{6.5}{*}{\begin{sideways}$n_{\rm qubits}=6$\end{sideways}}
  & 2 & 15 & 31 & \begin{tabular}{c} 2 \end{tabular} & \begin{tabular}{c} $3.68\cdot 10^{-2}$ \end{tabular} \\\cline{2-6}
  & 4 & [56, 80] & [84, 160] & \begin{tabular}{c} 2 \end{tabular} & \begin{tabular}{c} $2.31\cdot 10^{-4}$ \end{tabular} \\\cline{2-6}
  & 6 & [183, 300] & [255, 552] & \begin{tabular}{c} 2 \end{tabular} & \begin{tabular}{c} $2.93\cdot 10^{-5}$ \end{tabular} \\\cline{2-6}
  & 8 & [183, 300] & [255, 552] & \begin{tabular}{c} 2 \end{tabular} & \begin{tabular}{c} $4.53\cdot 10^{-5}$ \end{tabular} \\\cline{2-6}
  & 10 & [504, 888] & [682, 1568] & \begin{tabular}{c} 2 \end{tabular} & \begin{tabular}{c} $4.53\cdot 10^{-5}$ \end{tabular} \\
   \hline
\end{tabular}
\caption{For MPS-equivalent circuit with $n_{qubits}=4$ (top) and $n_{qubits}=6$ (bottom), the table displays the circuit details (bond dimension $D$, number of CNOT gates and number of single qubit gates), the number of optimization steps (i.e. $n_{\rm iters}$) and the KL divergence values. Since it remains unknown how many CNOT and single qubit gates are needed exactly to implement a general $k$-qubit unitary operator, we can only give lower and upper bounds (except for the case $D=2$).
Lastly, for $n_{\rm qubits}=4$, the upper bound for $D$ is 8: for $n_{qubits}=4$ and $D > 8$ the formula for the equivalent circuit breaks down. See Appendix \ref{appendix:qc_mps} for details about encoding MPS into quantum circuits.}
\label{table:MPS_depth_comparison}
\end{table}
\begin{table}[h!]
\centering
\begin{tabular}{||c|c c c c c||} 
 \hline
 \textbf{QCBM}& $n_{\rm layers}$ \hspace{5mm}& CNOT \hspace{5mm}&  1-qb \hspace{5mm}& $n_{\rm iters}$ \hspace{5mm}& KL \\ [0.5ex]
 \toprule
 \multirow{16}{*}{\begin{sideways}$n_{\rm qubits}=4$\end{sideways}} 
 &2&12&38& \begin{tabular}{c}10 \\100 \\ 1500 \end{tabular} & \begin{tabular}{c} $4.49\cdot 10^{-1}$\\ $1.16\cdot 10^{-1}$\\ $1.10\cdot 10^{-3}$ \end{tabular} \\\cline{2-6}
  & 4& 24 & 80 & \begin{tabular}{c}10 \\ 100 \\ 4000 \end{tabular} & \begin{tabular}{c} $6.78\cdot 10^{-1}$\\$1.28\cdot 10^{-1}$  \\ $1.82\cdot 10^{-7}$ \end{tabular} \\\cline{2-6}
  &6& 36 & 118 & \begin{tabular}{c}10\\ 100 \\ 4000 \end{tabular} & \begin{tabular}{c} $5.86\cdot 10^{-1}$\\ $2.95\cdot 10^{-1}$\\ $3.02\cdot 10^{-10}$ \end{tabular} \\\cline{2-6}
  &8& 48 & 156 & \begin{tabular}{c}10\\ 100 \\ 4000 \end{tabular} & \begin{tabular}{c} $6.50\cdot 10^{-1}$\\ $1.64\cdot 10^{-1}$ \\ $8.96\cdot 10^{-9}$ \end{tabular} \\\cline{2-6}
  &10& 60 & 194 & \begin{tabular}{c}10\\ 100 \\ 5000 \end{tabular} & \begin{tabular}{c} $5.43\cdot 10^{-1}$\\ $1.77\cdot 10^{-1}$  \\ $1.08\cdot 10^{-7}$ \end{tabular} \\
  \toprule
  \multirow{16}{*}{\begin{sideways}$n_{\rm qubits}=6$\end{sideways}}
  & 2 & 30 & 87 & \begin{tabular}{c}10\\ 1000 \\ 15000 \end{tabular} & \begin{tabular}{c} 1.088\\ $4.43\cdot 10^{-2}$ \\ $2.56\cdot 10^{-2}$ \end{tabular} \\\cline{2-6}
  & 4 & 60 & 180 & \begin{tabular}{c}10\\ 1000 \\ 30000 \end{tabular} & \begin{tabular}{c} 1.204\\ $5.81\cdot 10^{-2}$ \\ $1.43\cdot 10^{-3}$ \end{tabular} \\\cline{2-6}
  & 6 & 90 & 267 & \begin{tabular}{c}10\\ 1000 \\ 50000 \end{tabular} & \begin{tabular}{c} $9.82\cdot 10^{-1}$\\ $6.92\cdot 10^{-2}$ \\ $1.03\cdot 10^{-4}$\end{tabular} \\\cline{2-6}
  & 8 & 120 & 354 & \begin{tabular}{c} 10\\ 1000 \\ 15000\end{tabular} & \begin{tabular}{c} $8.81\cdot 10^{-1}$\\ $2.08\cdot 10^{-1}$ \\ $2.93\cdot 10^{-6}$\end{tabular} \\\cline{2-6}
  & 10 & 150 & 441 & \begin{tabular}{c}10\\ 1000 \\ 20000 \end{tabular} & \begin{tabular}{c} $8.69\cdot 10^{-1}$\\ $4.48\cdot 10^{-1}$ \\ $4.56\cdot 10^{-6}$\end{tabular}\\
   \hline
\end{tabular}
\caption{For QCBM with $n_{qubits}=4$ (top) and $n_{qubits}=6$ (bottom), the table displays the circuit details (number of layers, number of CNOT gates and number of single qubit gates), the number of optimization steps (i.e. $n_{\rm iters}$) and the KL divergence values.}
\label{table:QCBM_depth_comparison}
\end{table}

\subsection{Controlled rotations} \label{sec:crca_section}

\begin{figure*}
  \includegraphics[scale=0.45]{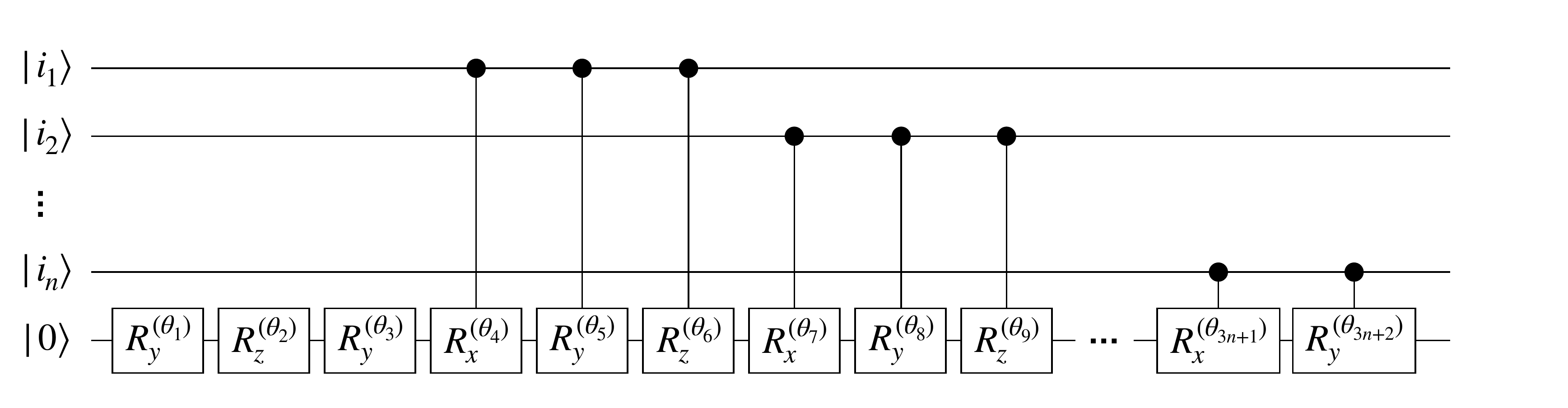}
  \caption{Controlled Rotation Circuit Ansatz (CRCA), the proposed near term circuit to implement operators of the form in Equation (\ref{VariationalFunctionOperator}). }
   \label{fig:CRCA}
\end{figure*}

In preparing the quantum state for the CVA problem we are faced with the problem of constructing operators that are all of the following form:
\begin{align}
    \label{FunctionOperator}
    R_f : \ket{i}\ket{0} \longmapsto \ket{i}\left( \sqrt{1- f(x_i)} \ket{0} + \sqrt{f(x_i}) \ket{1}\right), 
\end{align}
where the function $f: \Omega \longmapsto [0,1]$ and $x_i \in \Omega $ are discrete points chosen in its domain and lastly, the label $i \in \mathbb{N}$ is an integer indexing discrete points  whose binary expansion is $\sum_{k=0}^{k=n+m-1} 2^k i_k $ , with $i_k \in [0, 1]$.
If $f$ is efficiently computable classically with $\mathcal{O}({\rm poly}(n,m))$ resource, a common strategy for realizing $f$ exactly is by Reversible Logic Synthesis (\cite{d.m.millerd.maslovg.w.dueck2003, soekenm.dueckg.w.millerd.m.2016}); the cost of doing so would also be $\mathcal{O}\left({\rm poly}(m,n)\right)$.
The polynomial scaling is efficient in theory but in practice - and especially for near-term quantum computers - much more is to be desired in terms of lowering the circuit cost. For example, for the CVA instance considered in this work (with a total of 7 qubits), RLS requires a total of 89 CNOT gates. 
For the purpose of a near term alternative to RLS, we introduce an ansatz,\textit{ Controlled Rotation Circuit Ansatz} (CRCA) to try to approximate (\ref{FunctionOperator}) in the following ansatz (Figure \ref{fig:CRCA}):
\begin{equation}
    \label{VariationalFunctionOperator}
    \begin{array}{l}
    \widetilde{R}_f : \ket{i} \ket{0} \longmapsto \\ \quad \ket{i} \left( \sqrt{1 - \widetilde{f}(x_i, \vec{\theta})} \ket{0} + e^{i\phi(x_i, \vec{\theta})} \sqrt{\widetilde{f}(x_i, \vec{\theta})} \ket{1} \right)
    \end{array}
\end{equation}
where $\widetilde{f}$ is some function with the same domain and co-domain as $f$, and $\phi$ is some relative phase that depends on $x_i$ and $\vec{\theta}$. This ansatz can be used to approximate the operators $R_v$, $R_q$, and $R_p$ (see Equations \eqref{eq:R_E}, \eqref{eq:R_q}, \eqref{eq:R_p} in Section \ref{sec:Map_to_quantum}). Let $\widetilde{R}_v$, $\widetilde{R}_q$ and $\widetilde{R}_p$ denote approximations to each of the operators in the form described in Equation \eqref{VariationalFunctionOperator}. In the computational basis ${\rm span}\{|i\rangle\}\otimes{\rm span}\{|0\rangle,|1\rangle\}$ the operators $R_f$ and $\widetilde{R}_f$ can be arranged as block diagonal matrices with each block being an SU(2) rotation $U^{(i)}$ indexed by the label $i$:
\begin{eqnarray}
    \label{eq:Rf_block}
    R_f=
    \begin{pmatrix}
    U^{(1)} &          &        & \\
            & U^{(2)}  &        & \\
            &          & \ddots & \\
            &          &        & U^{(2^{n+m})}
    \end{pmatrix},\\
    U^{(i)}=
    \begin{pmatrix}
    \sqrt{1-f(x_i)} & -\sqrt{f(x_i)} \\
    \sqrt{f(x_i)} & \sqrt{1-f(x_i)}
    \end{pmatrix}.
\end{eqnarray}
For $\widetilde{R}_f$ we can similarly consider the block structure in \eqref{eq:Rf_block} with each block realizing the single-qubit rotation that leads to \eqref{VariationalFunctionOperator}. However, the relative phase factor $e^{i\phi(x_i,\vec{\theta})}$ is immaterial for the purpose of the CVA calculation, since ultimately the quantity desired $\langle\xi|\Pi|\xi\rangle$ (Section \ref{sec:Map_to_quantum}) is independent of the phase factors. We make a simplification and consider an operator $\widetilde{R}'_f$ which is equivalent to $\widetilde{R}_f$ for the purpose of CVA calculation but consists of only real elements:
\begin{eqnarray}
    \label{eq:Rpf_block}
    \widetilde{R}'_f=
    \begin{pmatrix}
    V^{(1)} &          &        & \\
            & V^{(2)}  &        & \\
            &          & \ddots & \\
            &          &        & V^{(2^{n+m})}
    \end{pmatrix},\\
    V^{(i)}=
    \begin{pmatrix}
    \sqrt{1-\widetilde{f}(x_i, \vec{\theta})} & -\sqrt{\widetilde{f}(x_i, \vec{\theta})} \\
    \sqrt{\widetilde{f}(x_i, \vec{\theta})} & \sqrt{1-\widetilde{f}(x_i, \vec{\theta})}
    \end{pmatrix}.
\end{eqnarray}
The error in CRCA training can then be characterized as the norm difference between the unitary operators $R_f$ and $\widetilde{R}'_f$. Based on the block structures in Equations \eqref{eq:Rf_block} and \eqref{eq:Rpf_block}, the norm difference becomes
\begin{equation}
    \epsilon_{{\rm CRCA},f}=\left\|R_f-\widetilde{R}'_f\right\|_2=\max_i\|U^{(i)}-V^{(i)}\|_2.
\end{equation}
Another common way to measure the approximation error of $\widetilde{f}$ with respect to $f$ is by the 1-norm difference:
\begin{equation}
    \label{ApproximateOperator1}
    \frac{1}{2^{n+m}}\sum_{i=0}^{{2^{n+m}-1}}|\widetilde{f}(x_i,\vec{\theta})-f(x_i)|.
\end{equation}
In Appendix \ref{sec:su2_compare} we show that $|\widetilde{f}(x_i,\vec{\theta})-f(x_i)|\le\|U^{(i)}-V^{(i)}\|{_2}$, which implies that $\epsilon_{{\rm CRCA}, f}$ is an upper bound to the 1-norm difference.

\begin{figure*}
    \centering
    \subfloat[Discount Factor]{\includegraphics[width=0.68\textwidth]{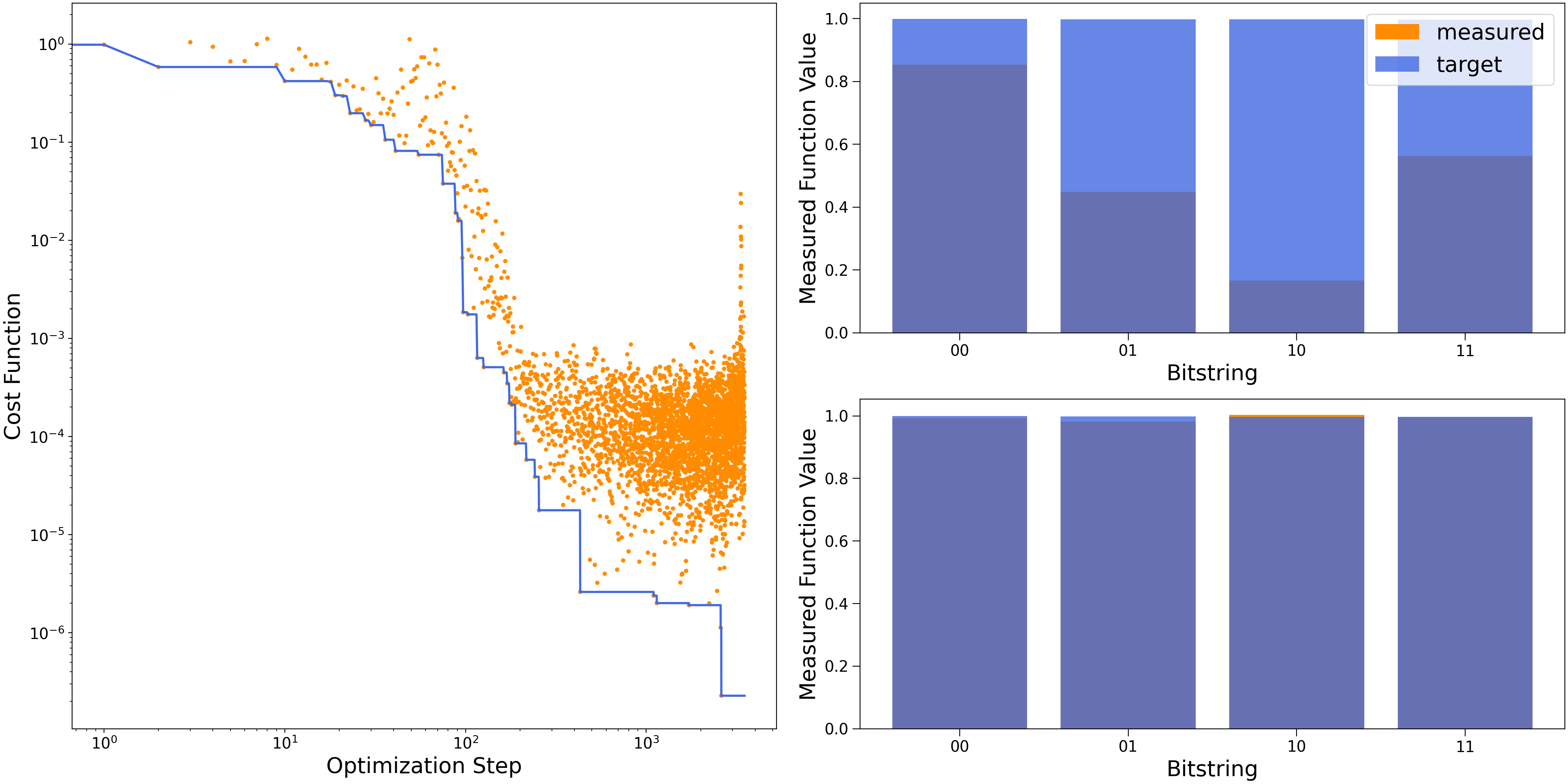}} 
    \vspace{2mm} 
    \quad
    \subfloat[Default probability]{ \includegraphics[width=0.68\textwidth]{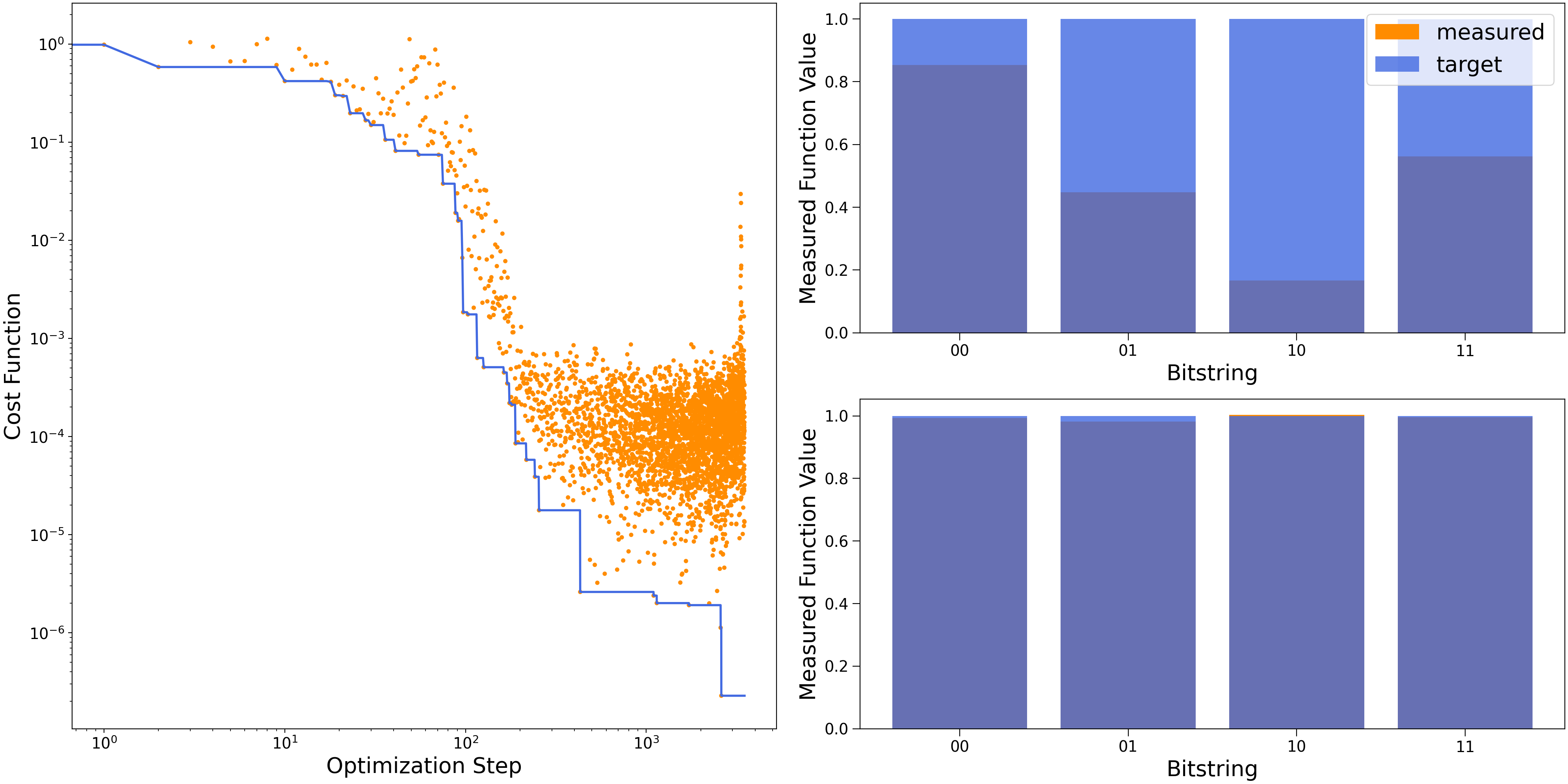}}
    \vspace{2mm} \\
    \quad
    \subfloat[Payoff function]{\includegraphics[width=0.68\textwidth]{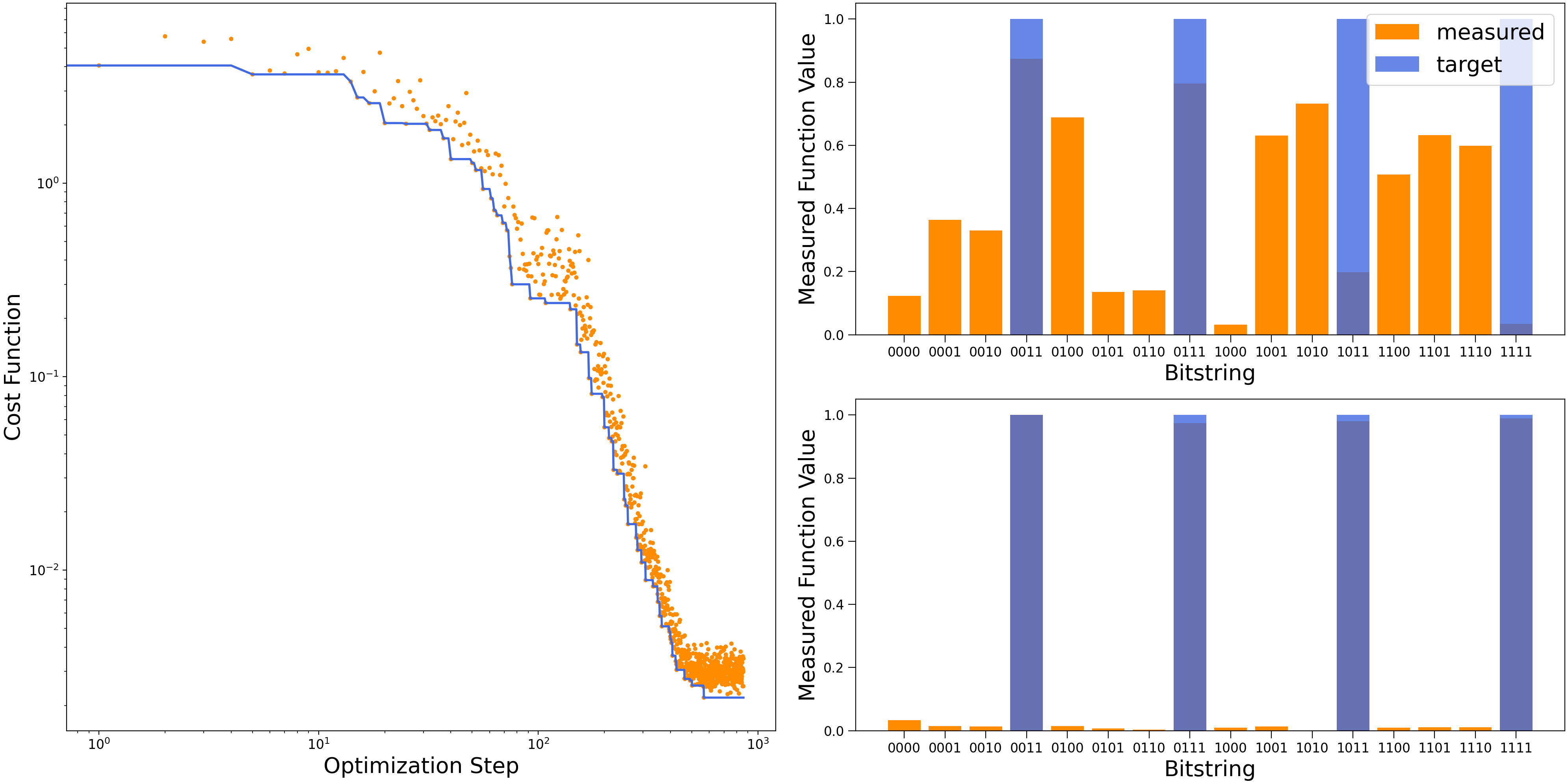}} 
    \hspace{6mm}
     \caption{The CVA circuit requires that we learn functions for \textbf{(a)} the discount factor $\tilde{p}(t_i)$, \textbf{(b)} the probability of default $\tilde{q}(t_i)$ and \textbf{(c)} the payoff function $\tilde{v}(s_j, t_i)$. For each subplot, on the right, the histograms display the functions encoded in the quantum state after the first training step and the one after the final step of training. The plots to the left display the value of the cost function in the training procedure.}
     \label{fig:single_asset_functions}
\end{figure*}

Having proposed CRCA as an alternative to RLS,  it is instructive to investigate the cost of RLS  not only to highlight the value of having CRCA, but also to look into what RLS produces in terms of the number of entangling operations (CNOT gates) for a practical problem. We use the transformation based synthesis method \cite{inproceedings} for our implementation of RLS. We assume the existence of three qubit registers, namely:
 
 \begin{enumerate}[label=(\roman*)]
     \item First register stores the domain values,
     \item Second register stores the function values,
     \item Third register contains the ancilla qubit for which the probability of measuring it in the $\ket{1}$ state gives the value of the function.
 \end{enumerate}
 
 More concretely, we have the following sequence of quantum operations:
 \begin{align}
      {U}_f:&  \ket{i}\ket{0}\ket{0} \longmapsto \ket{i}\ket{f(i)}\ket{0} \label{eq:rls_1},  \\ 
       {O}_f:& \ket{i}\ket{f(i)}\ket{0} \longmapsto \ket{i}\ket{f(i)} \left(a_i \ket{0} + b_i \ket{1} \right) \label{eq:rotations_2}, \\
      {U}_f^{\dagger}:& \ket{i}\ket{f(i)} \left(a_i \ket{0} + b_i \ket{1} \right) \longmapsto \ket{i}\ket{0}\left(a_i \ket{0} + b_i \ket{1} \right), \label{eq:rls_2}
 \end{align}
where $b_i$ is such that $|b_i|^2 = f(i)$. The quantum operation $U_f$ is implemented by RLS while $O_f$ is a quantum circuit that contains $\mathcal{O}(n)$ controlled rotations, where $n$ is the number of qubits in the first register.
Let $C$ be the number of CNOT gates from RLS, then we need a total of $2C$ CNOT gates to implement the evaluation of the function i.e (\ref{eq:rls_1}) and its un-computation i.e (\ref{eq:rls_2}). The number we report for $C$ will contain all the CNOT gates needed to implement functions for the payoff function, the discount factor, and default probabilities. For the step in (\ref{eq:rotations_2}) we require $n$ two qubit gates for both the discount factor and the default probabilities and $2n$ two qubit gates for the payoff function. 
Thus, the total number of two qubit gates $N_{2q}$ needed to solve a CVA problem instance is 
 \begin{equation}
     N_{2q} = 2C + n + n + 2n = 2 (C + 2n).
 \end{equation}
 In Table \ref{table:rls_results} we present numerical results for calculating $N_{2q}$ for different numbers of qubits. In comparison, to implement all the three desired functions for the CVA instance with CRCA, we need a total of 6 CNOTs for both the discount factor and the default probabilities and 24 CNOTs (2 layers) for the payoff function, giving a grand total of 36 CNOTs for a qubit size register of 4 qubits. Of course, while we expect to continue to see savings in the number of CNOT gates, the classical optimization problem for CRCA gets harder with the number of qubits.
 
\begin{table}[h!]
\centering
 \begin{tabular}{|| c c ||} 
 \hline
 \hspace{.5mm} $n$ $\quad$& $N_{2q}$ \hspace{.5mm}\\ [.5ex]
 \toprule
 \hspace{.5mm}4 $\quad$ & 212 \hspace{.5mm}\\ 
 \hline
 \hspace{.5mm}6 $\quad$& 640\hspace{.5mm} \\
 \hline
 \hspace{.5mm}8 $\quad$& 1428  \hspace{.5mm}\\
 \hline
 \hspace{.5mm}10 $\quad$& 25226 \hspace{.5mm} \\
 \hline
 \hspace{.5mm}12 $\quad$& 114632  \hspace{.5mm}\\
 \hline
 \hspace{.5mm}14 $\quad$& 483436  \hspace{.5mm}\\ [1ex]
 \hline
\end{tabular}
\caption{Cost of RLS. $n$ is the number of qubits, while $N_{2q}$ is the number of required two-qubit gates.}
\label{table:rls_results}
\end{table}

\subsection{Noiseless quantum CVA value}
\label{sec:quantum_cva}
We now have all the ingredients needed to build the quantum circuit shown in Figure \ref{fig:cva_circuit} so that we are able to get a quantum estimate for the CVA value. Once all the components are trained we can simply run the circuit and calculate the probability of three ancilla being in the state $\ket{111}$. Let $\ket{\widetilde{\xi}}$ be the actual output state of the quantum circuit (versus the ideal state $|\xi\rangle$ in Equation \ref{eq:ansatz_state}). The quantum CVA value is then: 
\begin{align}
\label{eq:CVA_Q}
    \widetilde{\rm CVA}_{Q} &= 2^m(1-R)C_p C_q C_v \bra{\widetilde{\xi}} \Pi \ket{\widetilde{\xi}} \\
    \label{eq:CVA_Q_value}
     &= 1.987 \cdot 10^{-5}
\end{align}
where the values of $C_p$, $C_v$, $m$, $C_q$  are reported in Table  \ref{table:circuit_params} and $R$ is reported in Table \ref{table:CVA_parameters}.
The above result is calculated from an exact simulation of the quantum circuit. The error in the CVA calculation can be decomposed as the following:
\begin{equation}
\begin{array}{l}
     |{\rm CVA}_{\rm MC}-\widetilde{\rm CVA}_{Q}| \le \\[0.1in]
     \qquad |{\rm CVA}_{\rm MC}-\widetilde{\rm CVA}(2)| + |\widetilde{\rm CVA}(2)-\widetilde{\rm CVA}_{Q}| \\
     \qquad = \epsilon_D + \epsilon_Q 
\end{array}
\end{equation}
where $\epsilon_D=|{\rm CVA}_{\rm MC}-\widetilde{\rm CVA}(2)|$ is the discretization error and $\epsilon_Q=|\widetilde{\rm CVA}(2)-\widetilde{\rm CVA}_{Q}|$ is defined as the error due to deviation of the trained quantum circuit from an ideal quantum circuit that prepares $|\xi\rangle$ in Equation \eqref{eq:ansatz_state}. With a slight abuse of notation regarding the operators forming the quantum circuit (Figure \ref{fig:cva_circuit}), we consider $|\xi\rangle=R_pR_qR_vG_\mathcal{P}|0^{n+m+3}\rangle$ and $|\widetilde{\xi}\rangle=\widetilde{R}_p\widetilde{R}_q\widetilde{R}_v\widetilde{G}_\mathcal{P}|0^{n+m+3}\rangle$ where $\widetilde{R}_p$, $\widetilde{R}_q$, $\widetilde{R}_v$, and $\widetilde{G}_{\mathcal{P}}$ denote the operators produced from training. The core component of understanding $\epsilon_Q$ is to bound the error in estimating $\Pi$, for which we have the following upper bound (derived in Appendix \ref{sec:error_bound}):
\begin{equation}
\label{eq:error_bound}
    \begin{array}{l}
    \epsilon_\Pi = |\langle\xi|\Pi|\xi\rangle-\langle\widetilde{\xi}|\Pi|\widetilde{\xi}\rangle|\le \\ 
    \quad\sqrt{2\cdot{\rm KL}(p_G||p_{tg})}+2(\epsilon_{{\rm CRCA},v}+\epsilon_{{\rm CRCA},q}+\epsilon_{{\rm CRCA},p})
    \end{array}
\end{equation}
where $\epsilon_{{\rm CRCA},f}=\|\widetilde{R}_f-R_f\|_2$. Here we use $p_G$ to denote the output probability distribution (Equation \ref{born_rule} and $p_A=|\Psi_A|^2$ for $\Psi_A$ in Equation \ref{eq:mps_Psi}) from the state preparation operator $G_\mathcal{P}$. Although slightly different objective functions are used for QCBM and MPS training, in both cases ($C(\vec{\theta})$ in Equation \ref{eq:log-likelihood} and $\mathcal{L}$ in Equation \ref{log_like}) they are related to the KL divergence between the generated distribution $p_G$ and the target distribution $p_{tg}$. 

In Table \ref{table:CRCA_training} we list the error quantities that are useful for gaining insight on the main sources error. Clearly, since the CVA instance contains only $n+m=4$ qubits, the dominant source of error is discretization $\epsilon_D$ (computed from Equations \ref{eq:CVA_MC} and \ref{eq:CVA_2}), compared with error in building the quantum circuit $\epsilon_Q$ (computed from Equations \ref{eq:CVA_2} and \ref{eq:CVA_Q_value}). However, as shown in Figure \ref{fig:cva1N}, $\epsilon_D$ can be quickly suppressed by increasing the number of qubits $n$ for encoding the asset price. Through Equation \eqref{eq:CVA_Q} we can obtain the error $\epsilon_\Pi=\epsilon_Q/(M(1-R)C_vC_pC_q)$ in estimating the observable $\Pi$ as listed in Table \ref{table:CRCA_training}. Based on the upper bound of $\epsilon_\Pi$, apparently the contribution from state preparation, which is $\sqrt{2\cdot{\rm KL}(p_G||p_{tg})}=0.0288$, dominates over the contributions from CRCA training. The value of $\epsilon_\Pi$ is well within the upper bound in \eqref{eq:error_bound}.

\begin{table}
\centering
\begin{tabular}{||c | c||}
\hline
 Discretization error $\epsilon_D$ & $4.376\cdot 10^{-5}$ \\
 \hline
 Noiseless quantum circuit error $\epsilon_Q$ & $7.638\cdot 10^{-6}$ \\
 \hline
 Noiseless observable error $\epsilon_\Pi$ & $8.836\cdot 10^{-3}$ \\
 \hline
 State preparation error KL$(p_G||p_{tg})$ & $4.150\cdot 10^{-4}$ \\
 \hline
 Payoff function error $\epsilon_{{\rm CRCA},v}$ & $3.218\cdot 10^{-3}$ \\
 \hline
 Discount factor error $\epsilon_{{\rm CRCA},q}$ & $1.545\cdot 10^{-4}$ \\
 \hline
 Default probability error $\epsilon_{{\rm CRCA},p}$ & $1.546\cdot 10^{-4}$ \\
 \hline
  \end{tabular} 
\caption{Error quantities of the quantum circuit for the CVA instance described in Table \ref{table:circuit_params}.}
\label{table:CRCA_training}
\end{table}

The analysis so far assumes perfect amplitude estimation, namely that we are able to obtain $\langle\widetilde{\xi}|\Pi|\widetilde{\xi}\rangle$ exactly. In reality, with both quantum amplitude estimation and classical Monte Carlo algorithms, there is always a statistical error $\epsilon_S$ due to the finite amount of computational resource, be it quantum circuit runs or Monte Carlo steps, being used for statistical estimation. In the next section, we discuss using a recently developed Bayesian amplitude estimation technique \cite{2006.09350} for performing the amplitude estimation. 
A central object of this algorithm is the \emph{engineered likelihood function} (ELF) which is used for carrying out Bayesian inference. See Appendix \ref{sec:ELF} for a detailed description of this algorithm. 

\section{Concrete resource estimation}
\label{sec:resource_estimation}
\begin{figure}
\includegraphics[width=\linewidth]{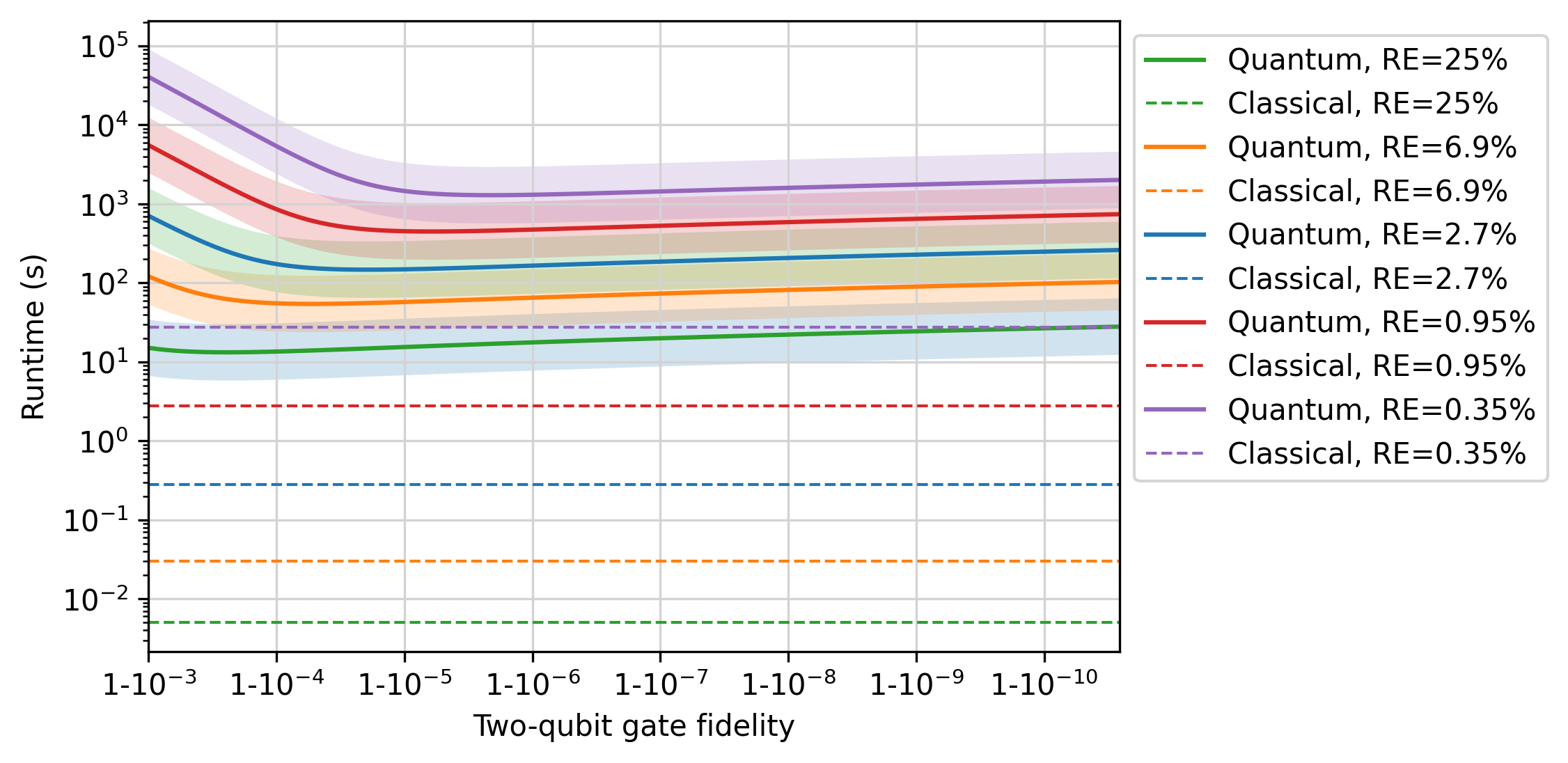}
\caption{Classical and quantum runtimes needed to estimate CVA to various accuracies. Here ``RE" is short for relative error. Each color represents a target accuracy, and the dashed line and solid curve of that color represent the classical and quantum runtimes needed to reach that accuracy, respectively. We assume that the quantum device uses the surface code in \cite{surfacecode2010} for quantum error correction and follow the method in \cite{huggins2020virtual} to analyze the overhead of this scheme. The surface code cycle time is set to $1\mu s$. }
\label{fig:runtime_estimates}
\end{figure}

In this section, we evaluate and compare the runtimes of classical and quantum algorithms for solving the CVA instance specified in Table \ref{table:CVA_parameters}.
For the classical benchmark, we compute the CVA value based on Equation \eqref{eq:CVA_exact}. In particular, to compute the expected exposure $E(t_i)$ at each time $t_i$, we run a classical Monte Carlo simulation with $10$, $100$, $1000$, $10000$ or $100000$ paths, respectively.
This leads to different runtimes of the algorithm and different errors in the outputs, which are illustrated in Figure \ref{fig:runtime_estimates}. 

For the quantum amplitude estimation, we apply the ELF-based amplitude estimation algorithm (Appendix \ref{sec:ELF}) to the circuit $A=\widetilde{R}_p\widetilde{R}_q\widetilde{R}_v\widetilde{G}_\mathcal{P}$ and the projection operator $\Pi$ given by Equation \eqref{eq:projector}. 
The ELF-based estimation scheme requires implementing two types of reflection operations:
\begin{enumerate}
    \item We implement the unitary operator $R_0(y)=\exp(iy\ketbra{0^{n+m+3}})$ with the method in \cite{Holmes_2020}, which requires $n+m+2$ ancilla qubits and $8(n+m+2)$ two-qubit gates \footnote{The idea of implementing $R_0(y)=\exp(iy\ketbra{0^n+m+2})$ is as follows. We first compute the logical AND of the bitwise NOT of the input by using $n+m+2$ ancilla qubits and $n+m+2$ Toffoli gates which are organized in a binary-tree fashion. Then we perform a $R_z$ rotation on the ancilla qubit that encodes the result. Finally, we undo the logical AND and bitwise NOT operations, restoring the ancilla qubits to their initial states. Each Toffoli gate can be decomposed into $4$ two-qubits gates, as implied by Theorem 8 of \cite{circuit_synthesis}.}.
    \item To implement the operator $U(x)=\exp(ix\Pi)$, we decompose it into $4$ two-qubit gates, as implied by Theorem 8 of \cite{circuit_synthesis}.
\end{enumerate}
  As a consequence, the quantum circuit for drawing samples that correspond to an ELF acts on $13$ logical qubits and contains $M=136$ logical two-qubit gates per layer. 

We assume that the quantum device has gate error rate $10^{-3}$ and uses the surface code in \cite{surfacecode2010} for quantum error correction, and we follow the method in \cite{huggins2020virtual} to analyze the overhead of this scheme. When the distance of the surface code is $d$, each logical qubit is mapped to $2d^2$ physical qubits, the fidelity of each non-Clifford gate is $f_{nc}\approx (1-10^{-(d+3)/2})^{100d}$, and the execution time of this gate is $t_{nc}\approx 100d \times$the surface code cycle time. Under a reasonable assumption about the layout of the circuit \cite{huggins2020virtual}, we have $f_{2Q}\approx f_{nc}$, and $t_{2Q} \approx t_{nc}$. We set the surface code cycle time to $1 \mu s$ (which is an optimistic estimate \cite{surfacecode2010}), and vary the code  distance from $10$ to $26$. This leads to different fidelities and execution times of logical two-qubit gates. We also assume the readout fidelity is $\bar{p}=f_{nc}^3$, as the projection operator $\Pi$ acts on three logical qubits. Under these assumptions about the hardware, we use the following equation to estimate the quantum runtimes needed to achieve the same error tolerance $\epsilon$ as in the classical experiments. 

\begin{widetext}
\begin{align}
    T_{\varepsilon} \sim O\left({t_{2Q}M} \cdot \dfrac{e^{-\lambda}}{\bar{p}^2}
    \left (\frac{ \lambda }{\varepsilon^2} + \frac{1}{\sqrt{2}\varepsilon} +\sqrt{\left(\frac{\lambda}{\varepsilon^2}\right)^2+\left(\frac{2\sqrt{2}}{\varepsilon}\right)^2}\right) \right ),
    \label{eq:runtime_model}    
\end{align}
\end{widetext}
where $t_{2Q}$ is the two-qubit gate time, $M$ is the number of two-qubit gates per layer \footnote{\cite{2006.09350} has assumed that the two-qubit gates are arranged in a bricklayer fashion and hence $M \approx nD/2$, where $n$ is the number of qubits and $D$ is the two-qubit gate depth per layer. This assumption does not hold here, as our ansatz circuit is highly sequential.}, $\lambda = M\operatorname{ln}(1/f_{2Q})$ in which $f_{2Q}$ is the two-qubit gate fidelity, and $\bar{p}$ is the read-out fidelity. In the low-noise limit, i.e.\ $\lambda \ll \epsilon$, Equation \eqref{eq:runtime_model} recovers the Heisenberg-limit scaling $O(1/\epsilon)$; while in the high-noise limit, i.e.\ $\lambda \gg \epsilon$, Equation \eqref{eq:runtime_model} recovers the shot-noise-limit scaling $O(1/\epsilon^2)$. Thus, this model interpolates between the two extreme cases as a function of $\lambda$. Such bounds allow us to make concrete statements about the extent of quantum speedup as a function of hardware specifications (e.g. the number of qubits and the two-qubit gate fidelity), and estimate runtimes using realistic parameters for current and future hardware.

\begin{figure}
\includegraphics[width=\linewidth]{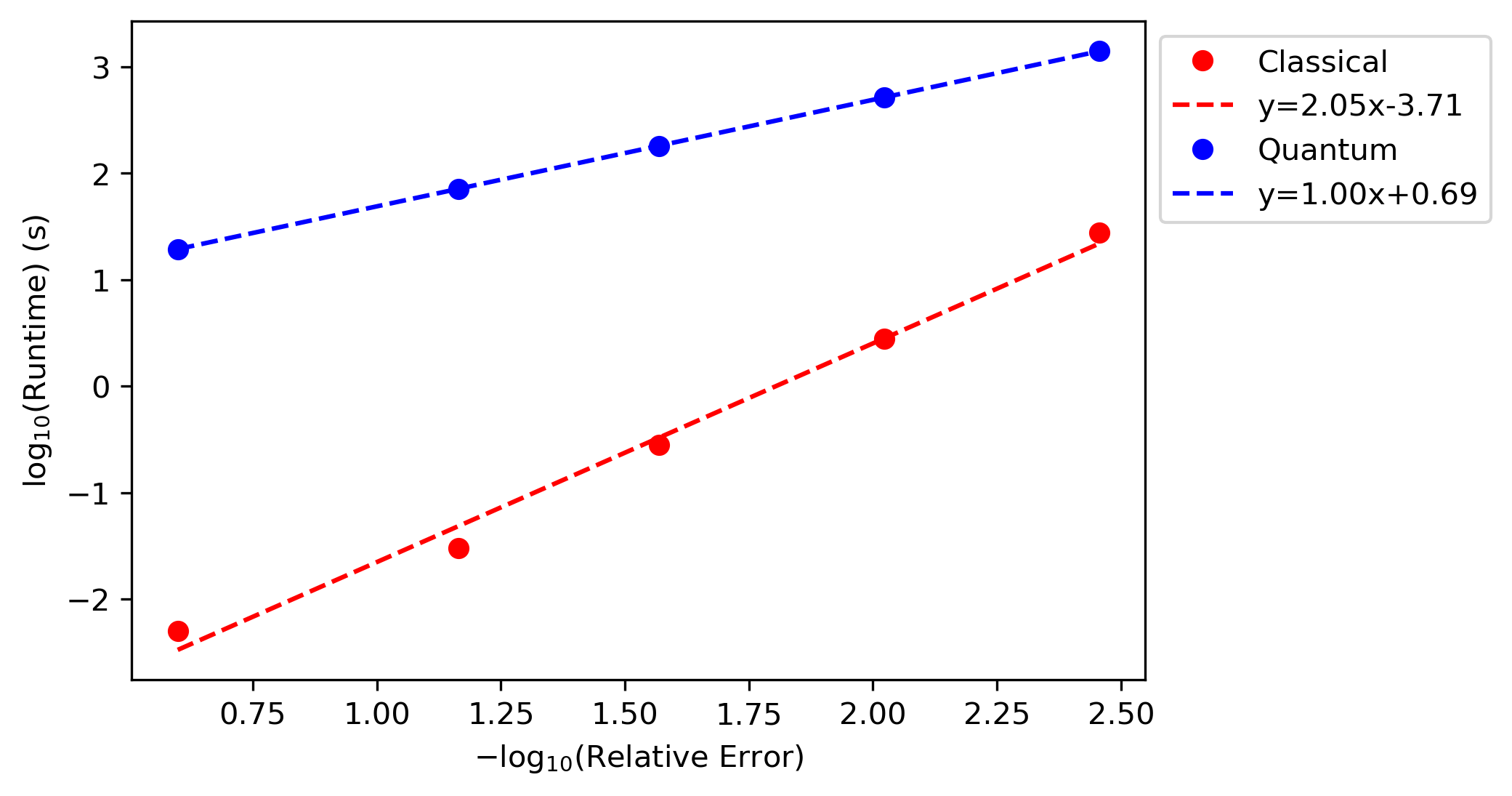}
\caption{Trends of classical and quantum runtimes for estimating CVA to higher and higher accuracies. Here we make the same assumption about the quantum device as in Figure \ref{fig:runtime_estimates}, setting the code distance to $18$. The classical runtime scales almost quadratically in the inverse error in the result, while the quantum runtime scales almost linearly in the same quantity.}
\label{fig:runtime_trend}
\end{figure}

The numerical results generated from the runtime model are demonstrated in Figure \ref{fig:runtime_estimates}. We observe that for estimating CVA to a relatively large error (e.g. $\ge 0.35\%$), our quantum algorithm runs slower than the classical algorithm. This is mainly because elementary quantum operations take much longer time to execute than their classical counterparts, due to the overhead from quantum error correction. So even though our algorithm contains fewer elementary operations than the classical algorithm, its runtime is still larger than the latter's. Nevertheless, as demonstrated by Figure \ref{fig:runtime_trend}, the runtime of classical algorithm scales almost quadratically in the inverse error in the result, while the runtime of our algorithm scales almost linearly in the same quantity, assuming the gate fidelity is sufficiently high. Therefore, as the desired accuracy of the result becomes higher, the gap between the quantum and classical runtimes will shrink, and eventually the quantum algorithm will surpass the classical one in efficiency. For example, based on the projection in Figure \ref{fig:runtime_trend}, our algorithm will run faster than the classical one for estimating CVA within relative error $\le 0.0067\%$. We emphasize that this threshold heavily depends on the hardware specifications, and could be dramatically shifted once we have better technology for realizing high-fidelity quantum gates. 

Our results are consistent with the ones in \cite{babbush2020focus} which show that it is unlikely to realize quantum advantage on a modest fault-tolerant quantum computer with quantum algorithms giving quadratic speedup over its classical alternatives, due to the large overheads associated with quantum error correction. For such quantum algorithms to run faster than their classical competitors, the input instance must be sufficiently large (as measured by the inverse accuracy of the result in this work), so that the small quantum scaling advantage can compensate for the large overhead factor from error correction. These findings suggest that we should either focus beyond quadratic speedups or dramatically improve the techniques for quantum error correction (or do both) in order to achieve quantum advantage on early generations of fault-tolerant quantum computers. 

While our observation of quantum advantage happening at large scales echoes the point in \cite{babbush2020focus}, our resource estimation is less pessimistic than the ones in that paper, due to the innovations in our circuit construction. Specifically, \cite{babbush2020focus} has focused on the case where both classical and quantum algorithms make calls to certain primitive circuits to solve the same problem, and the quantum primitive circuit is simply the coherent version of the classical one and is usually obtained via reversible logic synthesis. (Namely, the quantum algorithm has essentially the same spirit as the classical one but leverages amplitude amplification to achieve quadratic speedup.) Due to the overhead from reversible logic synthesis and quantum error correction, the quantum primitive circuit is much slower than its classical counterpart. However, this is not the case in this work, as we have invented alternative, shallower quantum circuits for CVA evaluation that are qualitatively different from the classical one based on Monte Carlo sampling, and consequently, our primitive calls take much less time than the ones considered in \cite{babbush2020focus}, which reduces the gap between the runtimes of classical and quantum algorithm at small scales.

\section{Discussion}\label{sec:discussion}

In this work we have described a general quantum algorithm for computing credit valuation adjustment and performed concrete resource estimation using a specific instance of the CVA problem along with the recently developed engineered likelihood function technique for quantum amplitude estimation. The study has revealed both challenges towards quantum advantage and opportunities for improving the quantum algorithm. Results from Section \ref{sec:resource_estimation} show that unless very small statistical error $\epsilon_S$ is desired (roughly on the order of $10^{-10}$) in estimating $\widetilde{\rm CVA}_Q$, the classical Monte Carlo implementation is faster than the quantum algorithm. Here we assume that $\epsilon_S$ is the dominant source of error, namely that all of the other error quantities in Table \ref{table:CRCA_training} are significantly smaller than $\epsilon_S$. Fulfilling this assumption requires further innovation on the quantum circuit construction. For instance, in order to suppress $\epsilon_D$ we need to increase the number of qubits encoding time and price parameters, which implies that scalable circuit generation training methods for both state preparation and control rotation are needed. The circuit generation methods are not only measured in terms of the scaling of their computational time but also the accuracy of the resulting circuit constructions, since this directly influences the $\epsilon_Q$ contribution. We have identified some of the recently developed tools \cite{FLIP,PECT,holmes2020efficient} that may help improving circuit generation, and active work is in progress to deploy and test those tools.

From a quantitative finance perspective, it is well-known that an European option pricing problem involving a single asset is analytically solvable, without the need for a Monte Carlo simulation. Nonetheless, our method can be generalized to other settings that do require stochastic simulation, such as multi-asset options, fat-tail distributions for $P(s,t)$, and stochastic intensity models \cite{stochastic-intensity}. As the landscape of risk analysis use cases in quantitative finance unfolds, we expect more innovations to be made in the future that push the frontier of quantum computing closer to practical advantage.

\section{Acknowledgements}
We thank Brian Dellabetta, Jerome Gonthier, Peter Johnson, Alejandro Perdomo-Ortiz, Max Radin for helpful discussions and support during the project.
We thank the team at BBVA for supporting the early parts of this study.
All of the numerical experiments in this work were carried out with Orquestra$^\text{\textregistered}$ \footnote{https://www.orquestra.io/} for workflow and data management.

\appendix
\section{QCBM Efficiency of Learning Discrete Distributions} \label{appendix:training_qcbm}
In Section \ref{subsec:QCBM}, we showed that a Quantum Circuit Born Machine defined on 4 qubits is able to learn the desired target distribution associated to the chosen problem instance. In order to perform the QCBM training, it is essential to collect a certain amount $M$ of measurements outcomes that build up the measured distribution to be compared to the target one. The QCBM training performance changes as the number of shots taken from the quantum computer varies. We performed an initial study to assess the efficiency of the algorithm in dependence on the number of quantum samples $M$ for different circuit sizes and depths. 
In order to quantify the QCBM efficiency, let's define the following measures: 
\begin{equation}
\begin{aligned}
& \epsilon_{\mu} = |\mu_{m}-\mu_{t}| \\
& \epsilon_{\sigma^2} = |\sigma^2_{m}-\sigma^2_{t}|,
\end{aligned}
\end{equation}
where $\mu$ and $\sigma^2$ are the mean and the variance of the measured ($m$) and the target ($t$) distribution. For simplicity, we limited the target distribution to one single log-normal peak. We computed the median of this quantity over 10 equivalent simulations for different value of $M$ and compared the scaling behaviour against the one obtained for a classical Monte Carlo simulation (i.e. rejection sampling). 

The relation between QCBM and Monte Carlo scaling depends on the circuit details. Specifically, if one increases the problem size while keeping the depth fixed, the QCBM efficiency decreases, whereas if one increases the depth keeping the number of qubits fixed, the QCBM efficiency increases and eventually outperforms its classical Monte Carlo counterpart. This initial analysis shows that the ``aspect ratio" of the quantum circuit matters when it comes to the efficiency of learning a discrete distribution. That is to say: given a $n$ qubits QCBM, it needs to have $f(n)$ layers in order to yield an advantage over Monte Carlo (see Table \ref{table:aspect-ratio}).
\begin{table}[h!]
\centering
 \begin{tabular}{| c || c c c c c |} 
 \hline
 $n$ & \quad2 \quad&  \quad3 \quad&  \quad4 \quad&  \quad5 \quad&  \quad6 \quad\\ 
 \hline
 $f(n)$&  \quad$<2$ \quad&  \quad$<2$ \quad&  \quad$\gtrsim4$ \quad&  \quad$\geq5$ \quad&  \quad$\geq7$ \quad\\
 \hline
\end{tabular}
\caption{Relation between number of qubits $n$ and number of layers $f(n)$ needed in the quantum circuit to yield advantage over the classical algorithm.}
\label{table:aspect-ratio}
\end{table}
 
\section{Quantum Circuits for Preparing Matrix Product States} 
\label{appendix:qc_mps}
Our method for encoding matrix product states into quantum circuits is based on the one in  \cite{PhysRevLett.95.110503}. Suppose $\ket{\psi}$ is an MPS given by
\begin{align}
	\ket{\psi}&=\bra{\psi_F} A^{[1]}  A^{[2]} \dots A^{[n]}  \ket{\psi_I}\\ &=\sum_{i_1, i_2, \dots, i_n \in \{0,1\}}\bra{\psi_F} A_{i_1}^{[1]}  A_{i_2}^{[2]} \dots A_{i_n}^{[n]} \ket{\psi_I}\ket{i_1 i_2…i_n},
\end{align}
where $A^{[j]}: \mathbb{C}^D \to \mathbb{C}^D \otimes \mathbb{C}^2$ satisfies $A^{[j]}=A_0^{[j]} \otimes \ket{0}+A_1^{[j]} \otimes \ket{1}$, for $j=1,2,\dots,n$, and $\ket{\psi_I}$, $\ket{\psi_F}  \in \mathbb{C}^D$ are arbitrary. Without loss of generality, we assume that $D=2^d$ for some $d \in \mathbb{Z}^+$ (if $D$ is not a power of two, we can embed this MPS into a larger MPS whose bond dimension is $2^{\ceil*{\operatorname{log}_2(D)}}$). We will find isometries $V^{[1]}$, $V^{[2]}$, $\dots$, $V^{[n]}$ such that 
\begin{align}
	\ket{\psi}= V^{[1]}  V^{[2]} \dots V^{[n]} \ket{\phi}
	\label{eq:mps_isometry}
\end{align}
for some state $\ket{\phi} \in \mathbb{C}^D$ by a sequence of singular value decompositions (SVDs). Specifically, we start by writing 
\begin{align}
\bra{\psi_F} A^{[1]} =V^{[1]}  M^{[1]},
\end{align}
where $V^{[1]}$ is the left unitary matrix in the SVD of the left-hand side, and $M^{[1]}$ is the remaining part of the SVD. Then we construct the other isometries by the following induction:
\begin{align}
(M^{[k]} \otimes I) A^{[k+1]} = V^{[k+1]} M^{[k+1]},	
\end{align}
where $V^{[k+1]}$ comes from the left unitary matrix in the SVD of the left-hand side, and $M^{[k+1]}$ is the associated remaining part of the SVD. After $n$ applications of SVDs from left to right, we set 
\begin{align}
\ket{\phi}=M^{[n]} \ket{\psi_I},
\end{align} 
obtaining Eq.~\eqref{eq:mps_isometry} as desired. It is easy to show that $V^{[k]}$ has dimension $\min(2D,2^k)\times \min(D,2^k)$. This implies that we can embed $V^{[k]}$ into a $2D$-dimensional or $2^k$-dimensional unitary $U^{[k]}$, depending on whether $k>d$ or not. (Precisely, we define $U^{[k]}$ as follows. If $k \le d$, then $U^{[k]}=V^{[k]}$; otherwise, $U^{[k]}$ can be any $2D$-dimensional unitary operator such that $U^{[k]} \ket{0} \ket{\eta} = V^{[k]} \ket{\eta}$ for all $\ket{\eta} \in \mathbb{C}^D$.) 

\begin{figure}
    \centering
    \includegraphics[width=0.9\linewidth]{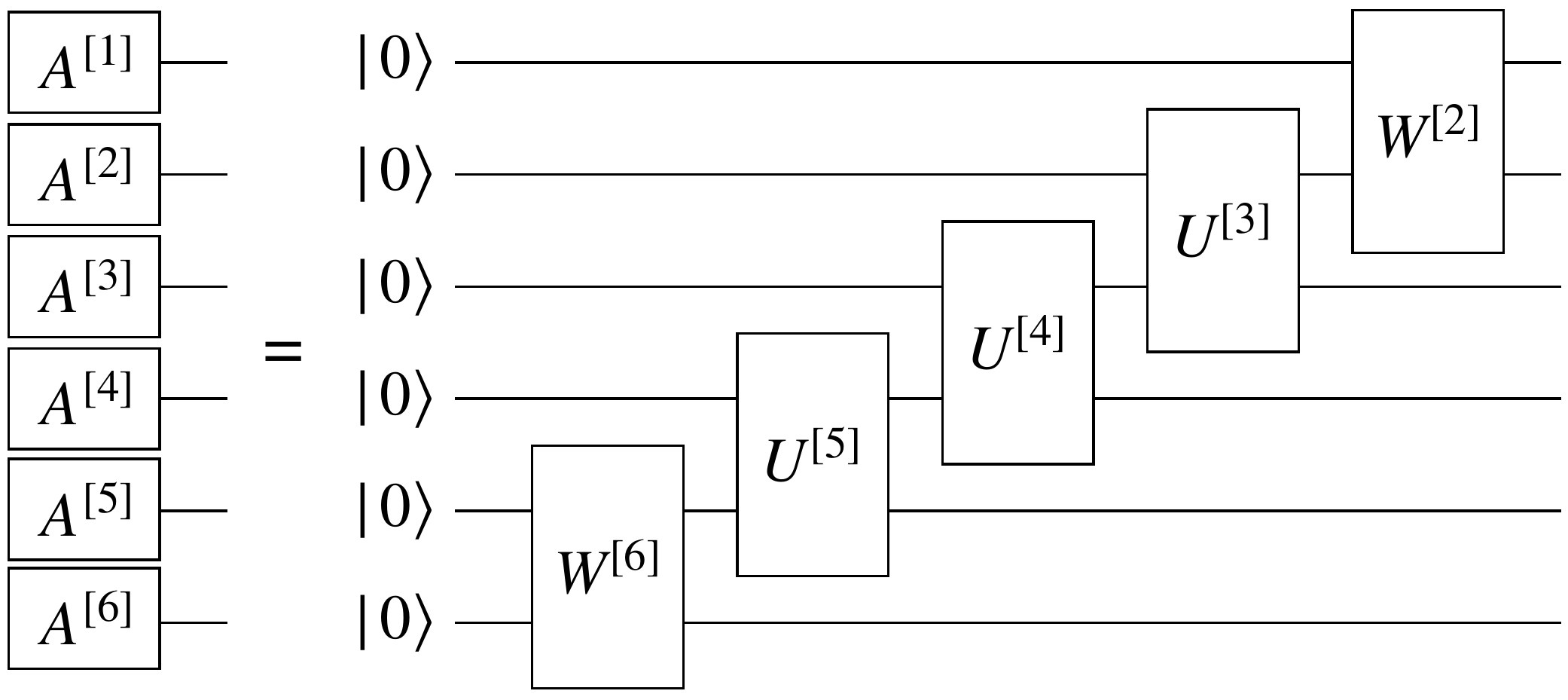}
    \caption{Quantum circuit for preparing a $6$-qubit MPS with bond dimension $2$. The two-qubit operators $W^{[2]}$, $U^{[3]}$, $U^{[4]}$, $U^{[5]}$ and $W^{[6]}$ are computed by the procedure in Appendix \ref{appendix:qc_mps}. }
    \label{fig:mps_circuit}
\end{figure}

Now we treat $\ket{\phi}$ as a $d$-qubit state (i.e. we use $d$ qubits to simulate the $D$-dimensional system). It follows that
\begin{align}
\ket{\psi} = U^{[1]} U^{[2]} \dots U^{[n]} \ket{0^{n-d}} \ket{\phi},     
\end{align}
where $U^{[k]}$ acts on the first $k$ qubits if $k \le d$, and the $(k-d)$-th to the $k$-th qubits otherwise. Let $Q$ be a $d$-qubit unitary operator such that $Q\ket{0^d} = \ket{\phi}$.
Then we get
\begin{align}
\ket{\psi} = W^{[d+1]} U^{[d+2]} \dots U^{[n-1]} W^{[n]} \ket{0^{n}},     
\label{eq:mps_qc}
\end{align}
where $W^{[d+1]}:=U^{[1]} U^{[2]} \dots U^{[d+1]}$ acts on the first $d+1$ qubits, and $W^{[n]}:=U^{[n]} (I \otimes Q)$ acts on the $(n-d)$-th to the $n$-th qubits. Eq. \eqref{eq:mps_qc} leads to a quantum circuit for preparing the state $\ket{\psi}$. Figure \ref{fig:mps_circuit} demonstrates this circuit for the case $n=6$ and $D=2$.

The unitary operators $W^{[d+1]}$, $U^{[d+2]}$, $\dots$, $U^{[n-1]}$ and $W^{[n]}$ can be decomposed into one- and two-qubit gates by using the method in \cite{circuit_synthesis}. Since each of these operators acts on $d+1$ qubits, it can be implemented by a circuit containing $c_{d+1}$ CNOT gates,  where 
$[4^{d+1}-3(d+1)-1]/4 \le c_{d+1} \le (23/48) \times 4^{d+1} - (3/2) \times 2^{d+1} + 4/3$. This implies that 
the number of CNOT gates in the final circuit is $\Theta(n4^d)=\Theta(nD^2)$. Table \ref{table:cnot_count_in_mps_qc} gives estimated CNOT counts for various $n$'s and $D$'s \footnote{Here we also give estimated numbers of single-qubit gates in the circuit for preparing a $n$-qubit MPS with bond dimension $D=2^d$. If arbitrary single-qubit gates can be used, then the number of single-qubit gates is $6n-5$ if $d=1$, or between 
$(n-d)(4^{d+1}-1)/3$ and $(n-d)(13\times 4^{d-1} - 3 \times 2^d)$ if $d\ge 2$. If only $R_x(\theta)$, $R_y(\theta)$ and $R_z(\theta)$ gates can be used, where $\theta \in \mathbb{R}$ is arbitrary, then the number of single-qubit gates is $12n-9$ if $d=1$, or between 
$(n-d)(4^{d+1}-1)$ and $(n-d)(21\times 4^{d-1} - 3 \times 2^d)$) if $d\ge 2$}.

\begin{table}[ht]
\centering
\begin{tabular}[t]{||lcccc||}
\hline
& D=2 & D=4 & D=8 & D=16\\
\toprule
n=4 & 9 & [28, 40] & -- & -- \\
n=8 & 21 & [84, 120] & [305, 500] & [1008, 1776] \\ 
n=16 & 45 & [196, 280] & [793, 1300] & [3024, 5328] \\
n=32 & 93 & [420, 600] & [1769, 2900] & [7056, 12432] \\
\hline
\end{tabular}
\caption{Estimated numbers of CNOT gates in the quantum circuits for MPS of various sizes and bond dimensions. Here $n$ is the number of qubits, and $D$ is the bond dimension. For each pair $(n, D)$, we give an upper bound and a lower bound on the corresponding CNOT count. Without loss of generality, we assume that $D \le 2^{\left \lfloor{n/2} \right \rfloor}$, since any $n$-qubit MPS can be transformed into an equivalent form with bond dimension at most $2^{\left \lfloor{n/2} \right \rfloor}$. So there are no entries for $(n=4, D=8)$ or $(n=4, D=16)$.}
\label{table:cnot_count_in_mps_qc}
\end{table}

\section{Comparing two SU(2) rotations for CRCA}
\label{sec:su2_compare}

We are concerned with relating the cost function of the form in Equation \eqref{ApproximateOperator1} to the norm difference $\|U-V\|_2$ between SU(2) rotations $U$ and $V$ such that
\begin{equation}
    U = 
    \begin{pmatrix}
    \cos\varphi & -\sin\varphi \\
    \sin\varphi & \cos\varphi
    \end{pmatrix},\quad
    V = 
    \begin{pmatrix}
    \cos\eta & -\sin\eta \\
    \sin\eta & \cos\eta
    \end{pmatrix}.
\end{equation}
It is clear that
\begin{equation}
    \begin{array}{ccl}
        \|U-V\|_2 & = & \|I-VU^\dagger\|_2 \\
         & = & \|I-\exp\left(-i(\varphi-\eta)\sigma_y\right)\|_2 \\
         & = & \displaystyle 2\left|\sin\left(\frac{\varphi-\eta}{2}\right)\right|.
    \end{array}
\end{equation}
\begin{figure}[t]
    \centering
    \includegraphics[scale=0.16]{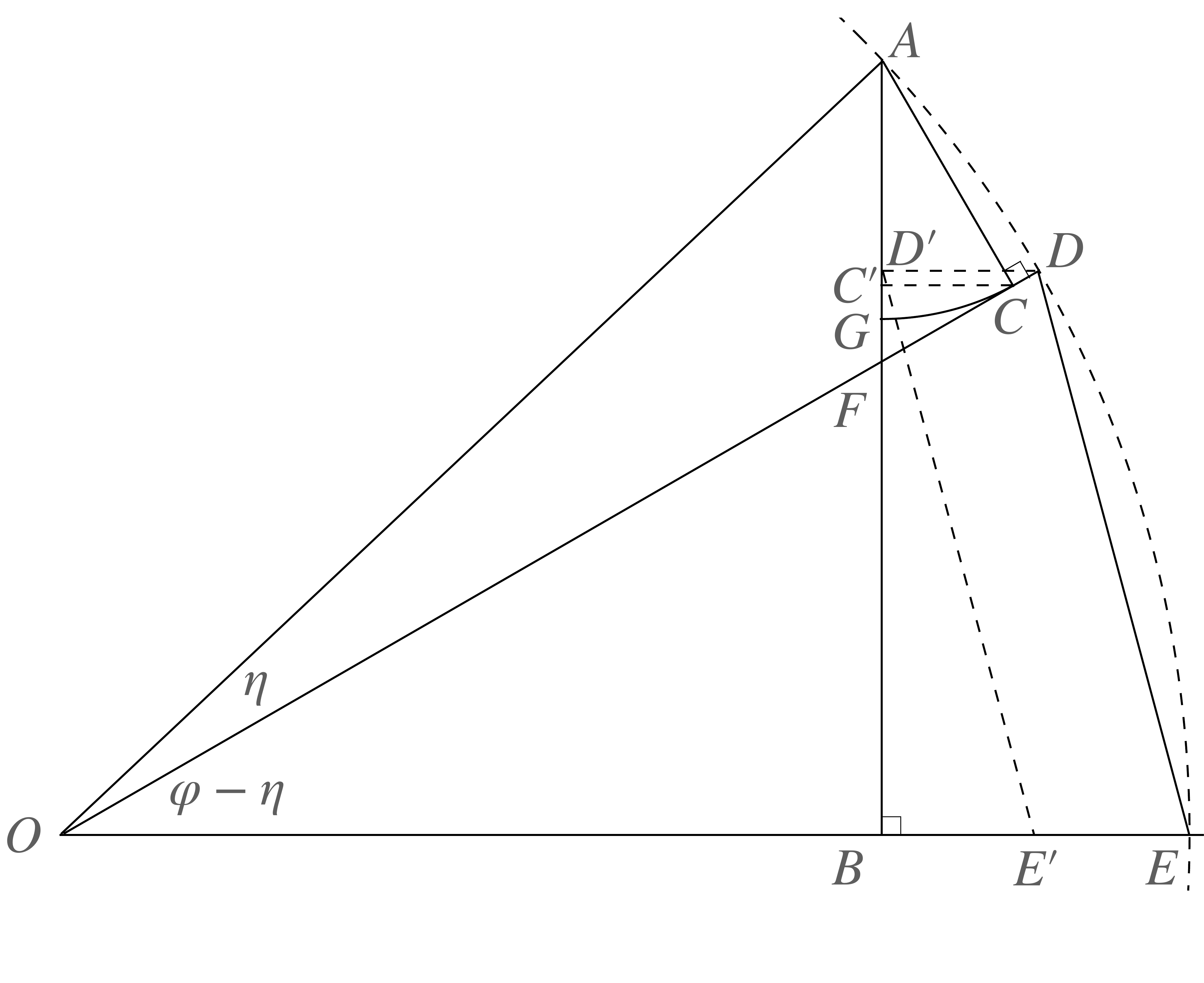}
    \caption{Geometric proof of inequality \eqref{eq:trig_ineq}. }
    \label{fig:trig_ineq}
\end{figure}
A quantity that is relevant to the training error of CRCA, as described in Equation \eqref{ApproximateOperator1}, is the difference $|\sin\varphi-\sin\eta|$. This quantity is equal to $|\tilde{f}(x_j,\vec{\theta})-f(x_j)|$ when $U$ is the SU(2) rotation in the block of the CRCA unitary indexed by $j$ and $V$ is the ideal rotation in the same block with all of its matrix elements being real numbers.
We show that 
\begin{equation}\label{eq:trig_ineq}
    |\sin\varphi - \sin\eta|\le 2\left|\sin\left(\frac{\varphi-\eta}{2}\right)\right|, 0\le\eta<\varphi\le\pi/2
\end{equation}
using a geometric illustration (Figure \ref{fig:trig_ineq}). Assume $0<\eta<\varphi<\pi/2$. From point $O$ we make three rays $\overrightarrow{OA}$, $\overrightarrow{OD}$ and $\overrightarrow{OE}$ such that $\angle AOD=\eta$ and $\angle DOE=\varphi-\eta$. Points $A$, $D$, and $E$ form an arc $\arc{ADE}$ on a unit circle. Therefore $|\overrightarrow{DE}|=2|\sin\left(\frac{\varphi-\eta}{2}\right)|$. From $A$ we make two lines $\overrightarrow{AC}$ and $\overrightarrow{AB}$ that are perpendicular to $\overrightarrow{OD}$ and $\overrightarrow{OE}$ respectively. Therefore $|\overrightarrow{AC}|=\sin\eta$ and $|\overrightarrow{AB}|=\sin\varphi$. Using $A$ as the center and $\overrightarrow{AC}$ as the radius, we make an arc $\arc{CG}$ which intercepts $\overrightarrow{AB}$ at $G$. Hence
\begin{equation}
\label{eq:BG}
    |\overrightarrow{BG}|=|\sin\varphi-\sin\eta|.
\end{equation}
From points $C$ and $D$ we make perpendicular lines to $|\overrightarrow{AB}|$ that intercepts $\overrightarrow{AB}$ at $C'$ and $D'$ respectively. Since $\eta>0$, $C$ is between $D$ and $F$. Since $\overrightarrow{CC'}$ is in parallel to $\overrightarrow{DD'}$, $C'$ is between $D'$ and $F$. Since $\angle C'CF=\angle DOE=\varphi-\eta$ and $\overrightarrow{CF}$ is a tangent of $\arc{CG}$, $G$ is between $F$ and $C'$, which is itself between $F$ and $D'$. Therefore we have that $G$ is between $F$ and $D'$, namely
\begin{equation}
\label{eq:ineq1}
    |\overrightarrow{BG}|<|\overrightarrow{BD'}|.
\end{equation}
From $D'$ we make a parallel line to $DE$ which intersects $\overrightarrow{AE}$ at $E'$. Then 
\begin{equation}
\label{eq:ineq2}
    |\overrightarrow{BD'}|<|\overrightarrow{E'D'}|=|\overrightarrow{ED}|=2\left|\sin\left(\frac{\varphi-\eta}{2}\right)\right|.
\end{equation}
Combining \eqref{eq:BG}, \eqref{eq:ineq1} and \eqref{eq:ineq2} yields the conclusion in \eqref{eq:trig_ineq}.

\section{Derivation of the CVA observable error bound}\label{sec:error_bound}

Here we derive the upper bound in \eqref{eq:error_bound}. We start from state preparation, where we generate $p_G(x_i)=|\langle x_i|G_\mathcal{P}|0^{n+m}\rangle|^2$ to approximate a target distribution $p_{tg}(x_i)$ for $x_i\in\Omega=\{0,1\}^{n+m}$, $i=1,\cdots,2^{n+m}$. We can then rewrite the error term as
\begin{widetext}
\begin{eqnarray}
    \displaystyle
    \left|\langle\xi|\Pi|\xi\rangle-\langle\widetilde{\xi}|\Pi|\widetilde{\xi}\rangle\right| 
    =\left|\sum_ip_{tg}(x_i)\langle x_i|R_v^\dagger R_q^\dagger R_p^\dagger\Pi R_pR_qR_v|x_i\rangle\right. 
    -\left.\sum_ip_G(x_i)\langle x_i|\widetilde{R}_v^\dagger\widetilde{R}_q^\dagger\widetilde{R}_p^\dagger\Pi\widetilde{R}_p\widetilde{R}_q\widetilde{R}_v|x_i\rangle\right| \\
    \displaystyle
    =\left|\sum_ip_{tg}(x_i)\langle x_i|R_v^\dagger R_q^\dagger R_p^\dagger\Pi R_pR_qR_v|x_i\rangle\right. 
    -\sum_ip_G(x_i)\langle x_i|R_v^\dagger R_q^\dagger R_p^\dagger\Pi R_pR_qR_v|x_i\rangle \nonumber \\
    \displaystyle
    +\sum_ip_G(x_i)\langle x_i|R_v^\dagger R_q^\dagger R_p^\dagger\Pi R_pR_qR_v|x_i\rangle 
    -\left.\sum_ip_G(x_i)\langle x_i|\widetilde{R}_v^\dagger\widetilde{R}_q^\dagger\widetilde{R}_p^\dagger\Pi\widetilde{R}_p\widetilde{R}_q\widetilde{R}_v|x_i\rangle\right| \\
    \displaystyle
    \le\sum_i\left|p_G(x_i)-p_{tg}(x_i)\right| 
    +\sum_ip_G(x_i)\left|\langle x_i|R_v^\dagger R_q^\dagger R_p^\dagger\Pi R_pR_qR_v|x_i\rangle\right. 
    -\left.\langle x_i|\widetilde{R}_v^\dagger\widetilde{R}_q^\dagger\widetilde{R}_p^\dagger\Pi\widetilde{R}_p\widetilde{R}_q\widetilde{R}_v|x_i\rangle|\right.
    \label{eq:err_bound1}
\end{eqnarray}
\end{widetext}

The first term in \eqref{eq:err_bound1} can be bounded from above by $\sqrt{2\cdot{\rm KL}(p_G||p_{tg})}$ due to Pinsker's inequality. The second term can be bounded from above by
\begin{eqnarray}
    \displaystyle
    \sum_ip_G(x_i)\left|\langle x_i|R_v^\dagger R_q^\dagger R_p^\dagger\Pi R_pR_qR_v|x_i\rangle\right. \makebox[1.1in]{} \nonumber \\
    \displaystyle
    -\langle x_i|\widetilde{R}_v^\dagger\widetilde{R}_q^\dagger\widetilde{R}_p^\dagger\Pi R_pR_qR_v|x_i\rangle \makebox[1.1in]{} \nonumber \\
    \displaystyle
    +\langle x_i|\widetilde{R}_v^\dagger\widetilde{R}_q^\dagger\widetilde{R}_p^\dagger\Pi R_pR_qR_v|x_i\rangle \makebox[1.1in]{} \nonumber \\
    \displaystyle
    -\left.\langle x_i|\widetilde{R}_v^\dagger\widetilde{R}_q^\dagger\widetilde{R}_p^\dagger\Pi\widetilde{R}_p\widetilde{R}_q\widetilde{R}_v|x_i\rangle|\right. \makebox[1.1in]{} \\
    \le\sum_ip_G(x_i)\left(\left|\langle x_i|\left(R_v^\dagger R_q^\dagger R_p^\dagger-\widetilde{R}_v^\dagger\widetilde{R}_q^\dagger\widetilde{R}_p^\dagger\right)\Pi R_pR_qR_v|x_i\rangle\right|\right. \nonumber \\
    +\left.\left|\langle x_i|\widetilde{R}_v^\dagger\widetilde{R}_q^\dagger\widetilde{R}_p^\dagger\Pi\left(R_pR_qR_v-\widetilde{R}_p\widetilde{R}_q\widetilde{R}_v\right)|x_i\rangle\right|\right) \makebox[0.3in]{} \label{eq:error_bound_inter}\\
    \le 2\left\|\widetilde{R}_p\widetilde{R}_q\widetilde{R}_v-R_pR_qR_v\right\|_2 \makebox[1.4in]{}
    \label{eq:error_bound2}
\end{eqnarray}
where going from \eqref{eq:error_bound_inter} to \eqref{eq:error_bound2} we apply the following argument:
\begin{equation}
    \begin{array}{l}
        \displaystyle
        \left|\langle 0|V^\dagger
        \underbrace{|111\rangle\langle 111|}_{\Pi}
        (V-U)|0\rangle\right|\\ 
        \makebox[0.4in]{}=\left|\langle 0|V^\dagger \underbrace{\sigma_x^{\otimes 3}|0\rangle}_{=|111\rangle}
        \right|\cdot \left|\langle 0|\sigma_x^{\otimes 3}(V-U)|0\rangle\right|\\
         \makebox[0.4in]{}\le \|V^\dagger \sigma_x^{\otimes 3}\|_2\cdot\|\sigma_x^{\otimes 3}(V-U)\|_2 \\
         \makebox[0.4in]{}=\|V-U\|_2.
    \end{array}
\end{equation}
When applying the argument for each term in the sum of \eqref{eq:error_bound_inter} we let $U$ and $V$ be unitary operators such that $U|0\rangle=\widetilde{R}_p\widetilde{R}_q\widetilde{R}_v|x_i\rangle$ and $V|0\rangle=R_pR_qR_v|x_i\rangle$. The operator $\sigma_x^{\otimes 3}$ acts on the three ancilla qubits that the projector $\Pi$ acts non-trivially on (Equation \ref{eq:projector}).

The term in \eqref{eq:error_bound2} can be further bounded from above by
\begin{eqnarray}
    \left\|\widetilde{R}_p\widetilde{R}_q\widetilde{R}_v-R_pR_qR_v\right\|_2 \makebox[1.5in]{} \nonumber \\
    \displaystyle
    =\left\|\widetilde{R}_p\widetilde{R}_q\widetilde{R}_v-\widetilde{R}_p\widetilde{R}_q{R}_v+\widetilde{R}_p\widetilde{R}_q{R}_v-R_pR_qR_v\right\|_2 \nonumber \\
    =\left\|\widetilde{R}_p\widetilde{R}_q\left(\widetilde{R}_v-{R}_v\right)+\left(\widetilde{R}_p\widetilde{R}_q-R_pR_q\right)R_v\right\|_2 \nonumber \\
    \le\left\|\widetilde{R}_v-{R}_v\right\|_2 + \left\|\widetilde{R}_p\widetilde{R}_q-R_pR_q\right\|_2 \makebox[0.7in]{} \label{eq:error_bound3} \\
    \le \cdots \makebox[2.5in]{} \nonumber \\
    \le \left\|\widetilde{R}_v-{R}_v\right\|_2 + \left\|\widetilde{R}_p-{R}_p\right\|_2 + \left\|\widetilde{R}_q-{R}_q\right\|_2
    \makebox[0.15in]{}
    \label{eq:error_bound4}
\end{eqnarray}
where in going from \eqref{eq:error_bound3} to \eqref{eq:error_bound4} we essentially repeat the same argument that leads to \eqref{eq:error_bound3}. Combining \eqref{eq:error_bound4} and \eqref{eq:err_bound1} yields \eqref{eq:error_bound}.

\section{Quantum amplitude estimation using engineered likelihood function (ELF)}
\label{sec:ELF}
Here we describe the quantum algorithm in \cite{2006.09350} for robust amplitude estimation. Suppose we want to estimate the expectation value
\begin{align}
\eta = \cosp{\theta} =\bra{A} O \ket{A},
\label{eq:defPi}
\end{align}
where $\ket{A}=A\ket{0^{k}}$ in which $A$ is a $k$-qubit unitary operator, $O=2\Pi-I$ in which $\Pi$ is a projection operator, and $\theta=\arccosp{\eta}$ is introduced to facilitate Bayesian inference later on. For the CVA problem, $A=\widetilde{R}_p\widetilde{R}_q\widetilde{R}_v\widetilde{G}_\mathcal{P}$ is the quantum circuit for preparing the state $\ket{A}=\ket{\widetilde{\xi}}$, and $\Pi$ is given by Equation \eqref{eq:projector}. Then $\bra{A} O \ket{A}=2 \widetilde{\text{CVA}}_Q/C - 1$, where $C=M(1-R)C_vC_pC_q$ by Equation \eqref{eq:CVA3}. So one can infer $\widetilde{\text{CVA}}_Q$ from the estimate of $\bra{A} O \ket{A}$.  

\begin{figure}
\center
\includegraphics[width=0.96\linewidth]{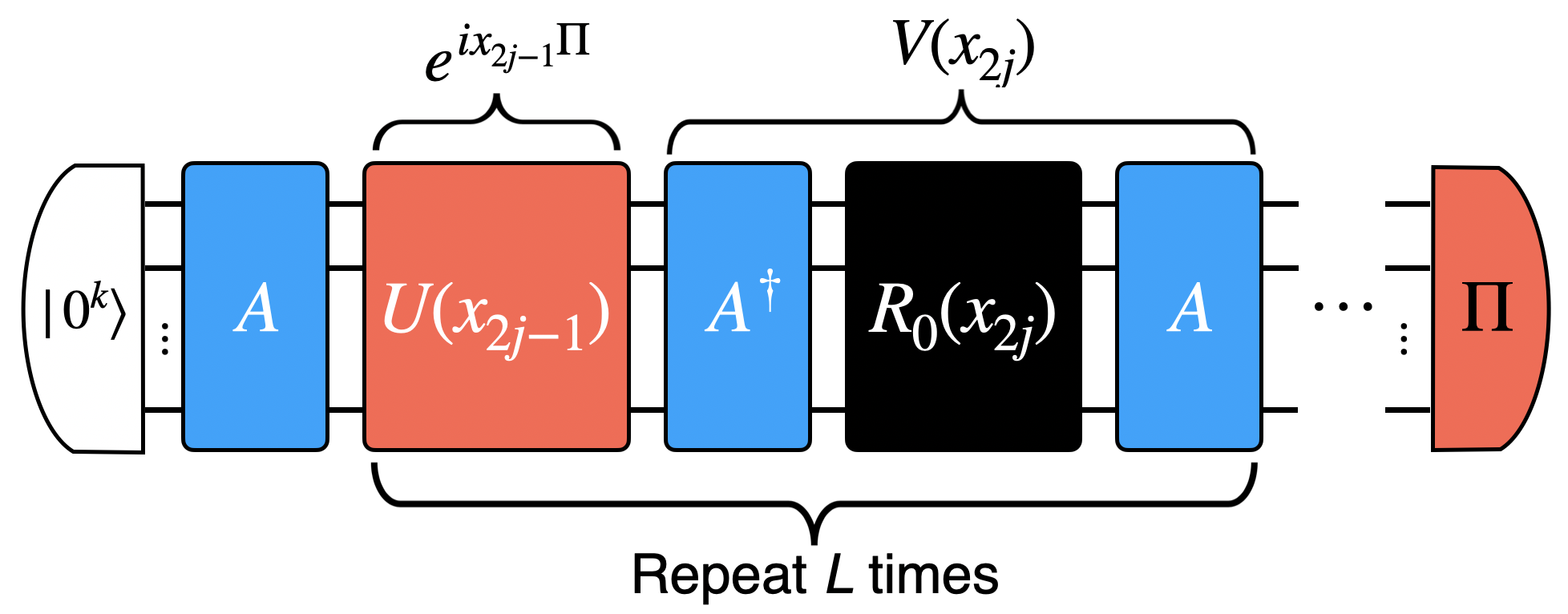}
\caption{Quantum circuit for generating samples that correspond to an engineered likelihood function. Here $R_0(y)=\exp(iy \ketbra{0^k})$ in which $y \in \mathbb{R}$ is arbitrary.}
\label{fig:circuit_diagram}
\end{figure}

We use the quantum circuit in Figure~\ref{fig:circuit_diagram} to generate the \emph{engineered likelihood function} (ELF), which is the probability distribution of a binary outcome $d \in \{0, 1\}$ given the unknown quantity $\theta$ to be estimated. The circuit consists of a sequence of unitary operations of forms $U(x)=\exp(i x \Pi)$ and $V(y)=A \exp(iy\ketbra{0^k}) A^{\dagger}$ in which $x, y \in \mathbb{R}$ are tunable parameters. Specifically, after preparing the ansatz state $\ket{A}=A\ket{0^k}$, we apply $2L$ unitary operations $U(x_1)$, $V(x_2)$, $\dots$, $U(x_{2L-1})$, $V(x_{2L})$ to it, varying the rotation angle $x_j$ in each operation. For convenience, we call $V(x_{2j})U(x_{2j-1})$ the $j$-th layer of the circuit, for $j=1, 2, \dots, L$. The output state of this circuit is 
\begin{align} 
Q(\vec{x})\ket{A}=V(x_{2L})U(x_{2L-1})\ldots V(x_2)U(x_1)\ket{A},
\end{align}
where $\vec x = (x_1,x_2,\ldots, x_{2L-1},x_{2L}) \in \mathbb{R}^{2L}$ contains the tunable parameters. Finally, we perform the projective measurement $\{\Pi, I-\Pi\}$ on this state, receiving outcome $d \in \{0,1\}$ with probability
\begin{align}
\mathbb{P}(d|\vec x)=\dfrac { 
1+(-1)^d \bra{A} Q^\dagger(\vec x) O Q(\vec x) \ket{A} }{2}.
\label{eq:lf}
\end{align}
This Bernoulli distribution depends on $\theta$ implicitly.   

In practice, the quantum circuit for generating the ELF is inevitably noisy, and the bias of the Bernoulli distribution in Eq.~\eqref{eq:lf} will be re-scaled by a factor of the fidelity of the circuit. Namely, if the fidelity of the circuit is $f \in [0, 1]$, then the probability of obtaining outcome $d \in \{0,1\}$ becomes
\begin{align}
\mathbb{P}(d|f, \vec x)=\dfrac { 
1+(-1)^d f \bra{A} Q^\dagger(\vec x) O Q(\vec x) \ket{A} }{2},
\label{eq:lf_noisy}
\end{align}
which still depends on $\theta$ implicitly.   

We use a Gaussian distribution to represent our knowledge of $\theta$ and keep updating this distribution until it is sufficiently concentrated around a single value. Specifically, we begin with an initial distribution of $\eta$ and convert it to the initial distribution of $\theta=\arccosp{\eta}$. Then we repeat the following procedure until convergence is reached. At each round, we first compute the circuit parameters $\vec x \in \mathbb{R}^{2L}$ that maximize the information gain from the measurement outcome $d$ (in certain sense). Then we run the quantum circuit in Figure \ref{fig:circuit_diagram} with the optimized parameters $\vec x$ and receive a measurement outcome $d \in \{0, 1\}$. After that, we update the distribution of $\theta$ by using Bayes' rule. Once this loop is finished, we convert the final distribution of $\theta$ to the final distribution of $\eta=\cosp{\theta}$, and set the mean of this distribution as the estimate of $\eta$. 

We have discovered that the efficiency of this algorithm is determined by the Fisher information of the engineered likelihood function at each round, and proposed efficient heuristic algorithms for finding the parameters $\vec x \in \mathbb{R}^{2L}$ that maximize this quantity. We have also found that the engineered likelihood function resembles a sinusoidal function in the critical region, and this fact allows us to perform Bayesian update efficiently without resorting to numerical integration. See \cite{2006.09350} for more details.

\bibliographystyle{unsrt}
\bibliography{references}

\end{document}